\documentclass[twocolumn]{aastex701}
\usepackage{xcolor}

\usepackage{amssymb}
\usepackage{tikz}
\usetikzlibrary{arrows.meta,positioning}
\usepackage{graphicx}

\usepackage{CJKutf8} 
\usepackage{amsmath,amssymb}   
\usepackage[shortlabels]{enumitem}
\usepackage{csvsimple,booktabs}

\newcommand{\Dysonian}{\mbox{Dysonian}}

\newcommand{\LsunNom}{\mathcal{L}_{\odot}}

\begin{document}

\title{WISE/CatWISE Constraints on Dysonian Waste-Heat Technosignatures in Nearby Galaxies}

\correspondingauthor{Tong-Jie Zhang, Zhen-Zhao Tao}
\email{tjzhang@bnu.edu.cn, taozhenzhao@dzu.edu.cn}

\author[orcid=0000-0002-8719-3137,sname='Huang']{Bo-Lun Huang}
\affiliation{Institute for Frontiers in Astronomy and Astrophysics, Beijing Normal University, Beijing 102206, China}
\affiliation{School of Physics and Astronomy, Beijing Normal University, Beijing 100875, China}
\email{Bolunh@hotmail.com} 
\author[orcid=0000-0002-4683-5500,sname='Tao']{Zhen-Zhao Tao}
\affiliation{College of Computer and Information Engineering, Dezhou University, Dezhou 253023, China}
\email{taozhenzhao@dzu.edu.cn}

\author[orcid=0000-0002-3363-9965,sname='Zhang']{Tong-Jie Zhang}
\affiliation{Institute for Frontiers in Astronomy and Astrophysics, Beijing Normal University, Beijing 102206, China}
\affiliation{School of Physics and Astronomy, Beijing Normal University, Beijing 100875, China}
\email{tjzhang@bnu.edu.cn}

\begin{abstract}
We search for galaxy-scale (\Dysonian) waste heat in the mid-infrared using \textit{WISE}. Starting from the 2MASS Redshift Survey (2MRS), we cross-match to \textit{CatWISE2020} and \textit{AllWISE}, apply standard MIR AGN/starburst vetoes (Stern, Assef R90, Jarrett), and treat W1 and W2 as stellar baselines and W3 and W4 as constraining bands. For each galaxy and for blackbody waste heat temperatures $T=150$--$600$\,K, we convert W3/W4 photometry into conservative $3\sigma$ per-galaxy upper limits on the bolometric waste heat luminosity using the \textit{WISE} bandpass (RSR) color correction. The resulting distributions have median caps of $\sim (5$--$9)\times 10^{8}\,L_{\odot}$ across $T=150$--$600$\,K. Aggregated at the \emph{population} level, the one-sided 95\% upper bound on the fraction of nearby galaxies that could host waste heat above a given threshold monotonically decreases with threshold and asymptotes to $\simeq 1/6500$ at high thresholds (set by the sample size). Sensitivity transitions from W4 at $T\!\lesssim\!200$\,K to W3 at $T\!\gtrsim\!300$\,K. Interpreted with the AGENT formalism, a fiducial Milky Way like stellar luminosity $L_\star\!=\!3\times 10^{10}\,L_{\odot}$ implies typical per galaxy caps of $\alpha \!\equiv\! L_{\rm wh}/L_\star \!\lesssim\!1.7$--$2.9\%$ over $T=150$--$600$\,K (e.g., $\alpha\!\lesssim\!1.8\%$ at $T{=}300$\,K). At $T\simeq300$\,K, no more than $f_{95}\simeq1.61\times10^{-4}$ ($\simeq0.0161\%$) of nearby galaxies can host KIII-scale systems reprocessing $\gtrsim21\%$ of a Milky Way--like stellar luminosity into $\sim300$\,K waste heat.
\end{abstract}

\section{Introduction}\label{sec:intro}

A basic consequence of energy use is the disposal of low entropy energy as heat \citep{Dyson1960}. If galaxy scale engineering reprocesses a significant fraction of starlight or other power sources, the associated waste heat should emerge as approximately thermal emission with a characteristic temperature of a few $\times 10^2$\,K, peaking in the mid-infrared (MIR). This motivates “\Dysonian\ SETI’’ \citep{Bradbury,C}, which seeks astrophysical waste heat signatures rather than deliberate beacons. \citet{Lacki2016} show that extremely cold ($\ll 100$\,K) galaxy scale “blackboxes’’ would violate microwave constraints and are therefore strongly disfavored. All-sky FIR/MIR searches began with \textit{IRAS}, with early constraints on Dyson-like structures presented by \citet{Carrigan2009}. With \textit{WISE} \citep{Wright2010} and \textit{NEOWISE} \citep{Mainzer2014}, \^G program \citep{Wright2014a,Wright2014b} set quantitative WISE-based upper limits on Kardashev Type~III systems; subsequent work tightly constrained the reddest extended sources \citep{Griffith2015}. 

Two developments now motivate a re–examination of galaxy–scale waste–heat constraints. First, the \textit{CatWISE2020} catalog \citep{Marocco2021}, built on WISE and the extended \textit{NEOWISE} time baseline, delivers improved W1/W2 astrometry and photometry across the entire sky, enabling cleaner baselines for modeling stellar continua in galaxies. Second, lessons from the \textit{WISE}/\textit{AllWISE} \citep{Cutri2013,AllWISEExplanatory} artifact taxonomy and the maturation of MIR AGN/starburst diagnostics (e.g., \citealt{Stern2012,Assef2018}; see also the widely used color wedges of \citealt{Jarrett2011}) allow us to construct transparent selection and masking policies that substantially reduce non–technological false positives.

In this paper we leverage \textit{CatWISE2020} (for W1/W2) together with \textit{AllWISE} (for W1,W2,W3,W4) to place \emph{population} constraints on galaxy–scale waste heat for a nearby, well–characterized parent sample. We adopt the 2MASS Redshift Survey (2MRS; \citealt{Huchra2012}) as our galaxy set and develop a fixed–configuration pipeline that:
\begin{enumerate}
\item cross–matches 2MRS to \textit{CatWISE2020} and \textit{AllWISE} with an explicit best–match protocol and per–band artifact vetting;
\item defines two quality policies, \textsc{Strict} and \textsc{Lenient}, differing only in the treatment of W4 artifacts, and applies three independent AGN/starburst masks (Stern, Assef~R90, and a Jarrett–style wedge) singly and in combination;
\item converts MIR magnitudes to fluxes (Vega zeropoints) and, for each galaxy and assumed radiator temperature $T\in\{150,200,300,400,600\}$\,K, derives a conservative per–object upper limit $L_{\rm wh}^{\max}(T)$ using a ``flux–cap'' approach that takes the most constraining of W3 or W4;
\item aggregates these per–object limits into \emph{population} upper bounds on the fraction of galaxies that could host waste–heat luminosity $\ge L_{\rm wh,thr}$ at temperature $T$, $f_{95}(T,L_{\rm wh,thr})\simeq 3/N_{\rm eff}$ (the $95\%$ upper limit for zero events).
\end{enumerate}
Our analysis yields (i) distributions of $L_{\rm wh}^{\max}$ across $T$ for the full sample, (ii) $f_{95}$ curves versus threshold that quantify survey sensitivity at the population level, and (iii) empirical diagnostics showing which MIR band (W3 or W4) actually sets the tightest constraint as a function of $T$. A comprehensive robustness check demonstrates that the main conclusions are insensitive to reasonable choices of artifact policy and mask combination.

This work therefore provides an updated, CatWISE–era assessment of how many K2/K3–like systems could be ``hiding in plain sight'' in the nearby Universe, under explicit assumptions documented in the text. The paper is organized as follows. Section~\ref{sec:data} describes the parent sample and MIR catalogs. Section~\ref{sec:methods} details the selection, masking, photometric conversions, and upper–limit methodology. Section~\ref{sec:results} presents per–object upper–limit distributions, population constraints, and band–of–limitation statistics, while Section~\ref{sec:robust} examines robustness to analysis choices. Section~\ref{sec:discussion} discusses implications and limitations, and Section~\ref{sec:conclusions} summarizes our conclusions. Throughout, we adopt Vega magnitudes for \textit{WISE} photometry and quote waste–heat luminosities in $L_\odot$.

\section{Data}\label{sec:data}

\subsection{Parent Sample: 2MRS}
We adopt the 2MASS Redshift Survey (2MRS) as our parent galaxy catalog \citep{Huchra2012}, which provides uniform all–sky positions and redshifts for nearby galaxies. For each source we retain the equatorial coordinates, redshift $z$, and luminosity distance $D_L$ as published by 2MRS.\footnote{All distances used in this work are the 2MRS luminosity distances; our conclusions do not depend on small variations in $D_L$ for the redshift range considered.} After basic sanity checks (valid coordinates and redshift), the working parent list contains $N\simeq2.24\times10^4$ galaxies.

\subsection{WISE Imaging and Catalogs}
We use mid–infrared photometry from the \textit{WISE} mission and its reactivation \citep{Wright2010,Mainzer2014}. Four bands are available: W1, W2, W3, and W4 (3.4, 4.6, 12, and 22\,$\mu$m). We combine two modern catalogs:

\begin{itemize}
\item \textbf{AllWISE} \citep{Cutri2013,AllWISEExplanatory} (dataset DOI: \doi{10.26131/IRSA1}), which supplies W1, W2, W3, W4 photometry, band–specific artifact flags (\texttt{cc\_flags}), photometric quality flags (\texttt{ph\_qual}), and extended–source information (\texttt{ext\_flg}). 
\item \textbf{CatWISE2020} \citep{Marocco2021} (dataset DOI: \doi{10.26131/IRSA551}), generated from unWISE coadds \citep{Lang2014,Meisner2017}, which provides improved W1/W2 astrometry and photometry and an unWISE artifact bitmask (\texttt{ab\_flags}).
\end{itemize}

All magnitudes are Vega. We adopt the AllWISE zero–magnitude flux densities $F_{\nu,0}=\{309.540,\,171.787,\,31.674,\,8.363\}$\,Jy for W1, W2, W3, W4, respectively (AllWISE Explanatory Supplement).\footnote{\url{https://wise2.ipac.caltech.edu/docs/release/allwise/expsup/}} For diagnostic purposes we also record per–band signal–to–noise (S/N) and reduced $\chi^2$ where available.

\subsection{Cross–matching and Best–Candidate Resolution}\label{sec:matching}
We perform a multi–object cone search of \textit{AllWISE} and \textit{CatWISE2020} at the 2MRS coordinates with a radius of $5\arcsec$.\footnote{The choice of $5\arcsec$ accommodates modest centroid differences for bright or extended galaxies while avoiding excessive false matches; tightening to $3\arcsec$ changes the sample negligibly after artifact vetting.} Because a given 2MRS position may return multiple WISE candidates (e.g., blended or nearby sources), we select a single ``best'' counterpart in each catalog per 2MRS target by ranking candidates with the following tie–breakers, in order: (1) fewer artifacts (CatWISE: \texttt{ab\_flags}==0 preferred; both catalogs: per–band \texttt{cc\_flags} clean), (2) smaller angular separation from the input position, and (3) stronger W1 profile–fit quality (higher S/N or smaller $\sigma_{\rm m}$). This yields one AllWISE row and one CatWISE row (or none) per 2MRS galaxy. Final sample construction is handled in Section~\ref{sec:methods}.

\subsection{Sample Summary}\label{sec:samplesummary}
\begin{figure}
  \centering
  \includegraphics[width=0.75\linewidth]{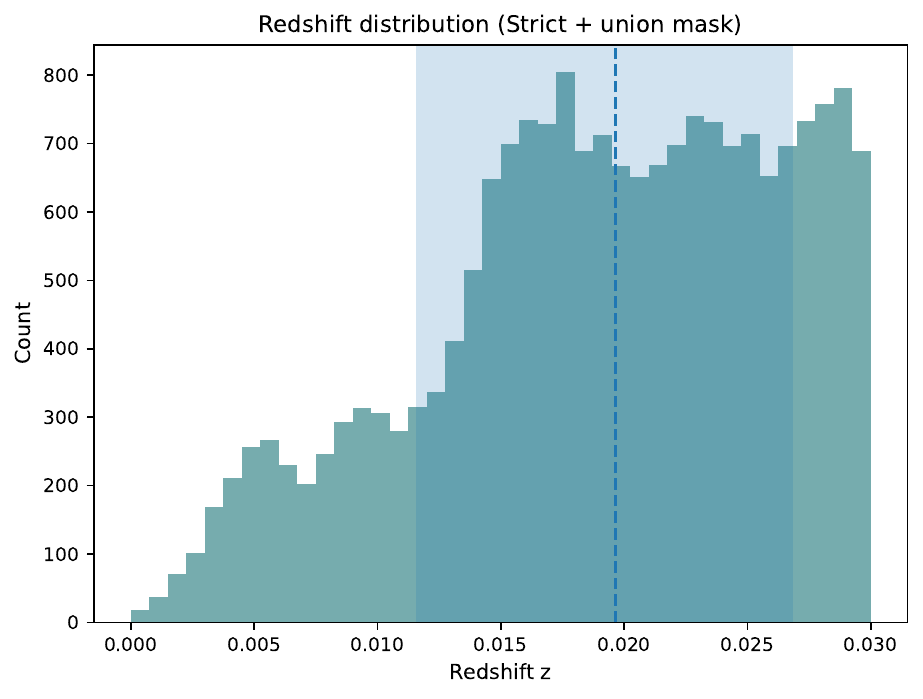}
  \caption{Redshift distribution of the working sample (Strict + union masks). The median and central 68\% interval are indicated.}
  \label{fig:zdist}
\end{figure}

Figure~\ref{fig:zdist} shows the redshift distribution for the final working sample used in the analysis (see Methods\,\S\ref{sec:policy} and \S\ref{sec:masks} for the definitions of the artifact policy and AGN/starburst masks). The median and central 68\% interval are $z_{\rm med}=0.01966$ with $[z_{16},z_{84}]=[0.01155,\,0.02683]$. The total number of galaxies with valid per object limits (defined after all selections in \S\ref{sec:results-sample}) is $N_{\rm valid}=19{,}617$, consistent with our nearby galaxy focus.

\section{Methods}\label{sec:methods}
Our analysis converts WISE photometry and flags into \emph{per–galaxy} waste–heat upper limits and then into \emph{population–level} bounds. For transparency, each step is specified by explicit equations and a short pseudo code workflow (Appendix~\ref{app:workflow}). All configuration choices (artifact policy, AGN/starburst masks, temperature grid, and significance level) are fixed before inspecting the population results.

\subsection{Band Choice and Artifact Policies}\label{sec:policy}
We treat W1/W2 as the stellar baseline and W3/W4 as the constraining bands for waste heat. Band selection follows:

\begin{itemize}
\item \textbf{W1/W2 (baseline):} prefer CatWISE magnitudes when \texttt{ab\_flags}=0 \emph{and} the per–band character of \texttt{cc\_flags} is ``0'' (clean). If CatWISE fails these criteria, we fall back to AllWISE provided its \texttt{cc\_flags} is clean in that band.
\item \textbf{W3/W4 (constraints):} use AllWISE.
\end{itemize}

We define two artifact policies that differ only in the treatment of W4; in all cases W1, W2, and W3 must be clean in their respective bands:
\begin{description}
\item[STRICT] W1, W2, and W3 require \texttt{cc\_flags} character “0” (clean); W4 must also have \texttt{cc\_flags} “0”.
\item[LENIENT] Identical to STRICT for W1, W2 and W3; W4 is retained even if its \texttt{cc\_flags} character is non‑zero (retained with a flag).
\end{description}

\subsection{AGN/Starburst Masks}\label{sec:masks}
We remove sources consistent with strong AGN or starburst MIR colors using three independent diagnostics applied to the WISE selection above:

\begin{enumerate}
\item \textbf{Stern cut:} $(W1{-}W2)\ge 0.8$ for bright $W2$; we adopt $W2<15.05$ as in \citet{Stern2012}.\\

\item \textbf{Assef R90:} $(W1{-}W2)\ge \mathcal{R}_{90}(W2)$, where $\mathcal{R}_{90}(W2)$ is the magnitude–dependent reliability–optimized color cut given by \citet[]{Assef2018}. For reference, \citet[]{Assef2018} also provide the completeness–optimized C75 relation; we use R90 in the analysis.\\

\item \textbf{Jarrett wedge:} $(W2{-}W3)\ge 1.3$, $(W1{-}W2)\le 1.7$, and $(W1{-}W2)\ge 0.10\,(W2{-}W3)+0.38$ and $\ge 0.315\,(W2{-}W3)-0.222$.
\end{enumerate}

Our analysis uses the union mask (Stern+Assef+Jarrett). We also evaluate variants (none; Stern only; Stern+Assef) as robustness checks (Fig.~\ref{fig:f95-robust}).

A KIII-like MIR excess moves sources toward large $W2{-}W3$ and moderately large $W1{-}W2$, overlapping the starburst/AGN locus. If such systems exist, they would be removed by our union mask \emph{when we require an \emph{astrophysically clean} retained sample}. However, our \emph{population} limits are set by the number of galaxies for which our data would have been sensitive at a given threshold ($N_{\rm eff}$). Figure~\ref{fig:f95-robust} shows that using \emph{no mask}, \emph{Stern only}, \emph{Stern+Assef}, or the full union yields statistically indistinguishable $f_{95}$ curves: the asymptotic plateaus and mid‑threshold behavior are unchanged within our precision.

\subsection{Photometric Conversions}\label{sec:phot}
For a Vega magnitude $m_b$ in band $b$, the flux density is
\begin{equation}
F_{\nu,b} \;=\; F_{\nu,0,b}\;10^{-0.4\,m_b},
\end{equation}
with $F_{\nu,0,b}$ taken from the WISE/AllWISE documentation. Magnitude errors are propagated as
\begin{equation}
\sigma_{F_{\nu,b}} \;\simeq\; F_{\nu,b}\,\left(\frac{\ln 10}{2.5}\right)\sigma_{m_b}.
\end{equation}
When \texttt{ph\_qual}=U, the catalog magnitude is an upper limit and we use only the cap defined below.

\subsection{Per–Galaxy Waste–Heat Upper Limits}\label{sec:pergal}
We assume that any waste–heat component from technological activity radiates approximately as a blackbody of temperature $T$, and we use W3/W4 to set a conservative “flux–cap’’ on its allowed normalization. For each band $b\in\{\mathrm{W3},\mathrm{W4}\}$ we define a cap that treats detections and non–detections at a common $k\sigma$ significance,
\begin{equation}
F_{\nu,b}^{\rm cap} =
\begin{cases}
\max\!\bigl[\,0,\;F_{\nu,b}+k\,\sigma_{F_{\nu,b}}-F_{\nu,b}^{\rm RJ}\,\bigr], & \text{(detection)}\\[4pt]
\max\!\bigl[\,0,\;\alpha_U\,F^{(U)}_{\nu,b}-F_{\nu,b}^{\rm RJ}\,\bigr], & \text{(\texttt{ph\_qual}=U),}
\end{cases}
\label{eq:fluxcap}
\end{equation}
where $k=3$, $F^{(U)}_{\nu,b}$ is the catalog upper limit, and $\alpha_U\equiv k/k_U$ rescales it to the same significance as detections (we adopt $k_U\simeq 2$; varying $k_U$ in $[1.5,3]$ affects only the faint end). The term $F_{\nu,b}^{\rm RJ}$ represents a possible Rayleigh–Jeans extrapolation of the stellar continuum from W1/W2; in the analysis we set $F_{\nu,b}^{\rm RJ}=0$ (i.e., we do not subtract any stellar contribution in W3/W4), and we assess the impact of enabling the RJ term as a robustness check in §\ref{sec:robust-rj}.

To map a single–band cap to a bolometric cap we use the \textit{WISE} \emph{relative spectral response} (RSR) and the band–averaged Planck function,
\begin{equation}
\left\langle B_\nu(T)\right\rangle_b \;\equiv\;
\frac{\displaystyle \int R_b(\nu)\,B_\nu(T)\,d\nu}{\displaystyle \int R_b(\nu)\,d\nu},
\label{eq:bbavg}
\end{equation}
where $R_b(\nu)$ is the RSR curve for band $b$ and $B_\nu(T)$ is the Planck function.

We compute $\langle B_\nu\rangle_b$ by integrating $R_b(\nu)B_\nu(T)$ in frequency space. The bolometric factor uses $\int_0^\infty B_\nu(T)\,d\nu=\sigma T^4/\pi$, so Eq.~\ref{eq:fbol} becomes 
$F_{\rm bol}^{(b)}(T)=F^{\rm cap}_{\nu,b}\,[\sigma T^4/\pi]/\langle B_\nu(T)\rangle_b\times 10^{-26}$. 
We also quote the WISE pivot wavelengths ($\lambda_{\rm W3}=11.56\,\mu$m, $\lambda_{\rm W4}=22.09\,\mu$m), defined by $\lambda_{\rm piv}^2=\frac{\int R(\lambda)\lambda\,d\lambda}{\int R(\lambda)\,d\lambda/\lambda}$, for monochromatic comparisons.

In practice we therefore write
\begin{equation}
F_{\rm bol}^{(b)}(T) \;=\; F_{\nu,b}^{\rm cap}\;
\frac{\displaystyle \int_0^\infty B_\nu(T)\,d\nu}
{\displaystyle \left\langle B_\nu(T)\right\rangle_b}\times 10^{-26}\quad [\mathrm{W\,m^{-2}}].
\label{eq:fbol}
\end{equation}

The per–band bolometric caps are then converted to luminosity caps via
\begin{equation}
L_{\rm wh}^{(b)}(T) \;=\; 4\pi D_L^2\,F_{\rm bol}^{(b)}(T),\qquad
L_{\rm wh}^{\max}(T) \;=\; \min\!\Bigl[L_{\rm wh}^{(\mathrm{W3})}(T),\,L_{\rm wh}^{(\mathrm{W4})}(T)\Bigr],
\label{eq:lwh}
\end{equation}
where $D_L$ is the luminosity distance. For all luminosity conversions we adopt the IAU nominal solar luminosity $\LsunNom = 3.828\times10^{26}\,\mathrm{W}$ \citep{Prsa}. We also record which band is limiting (Fig.~\ref{fig:which-band}). We evaluate $L_{\rm wh}^{\max}$ on a temperature grid $T=\{150,200,300,400,600\}$~K. We find that the RSR treatment increases $L_{\rm wh}^{\max}$ by $\sim$0–1\% at $T=150$–200~K, $\sim$9\% at $T=400$~K, and $\sim$16\% at $T=300$~K, while leaving the high–threshold population plateau unchanged.

\subsection{Population Bounds}\label{sec:pop}
\begin{figure*}[t!]
  \centering

  \includegraphics[width=0.6\textwidth]{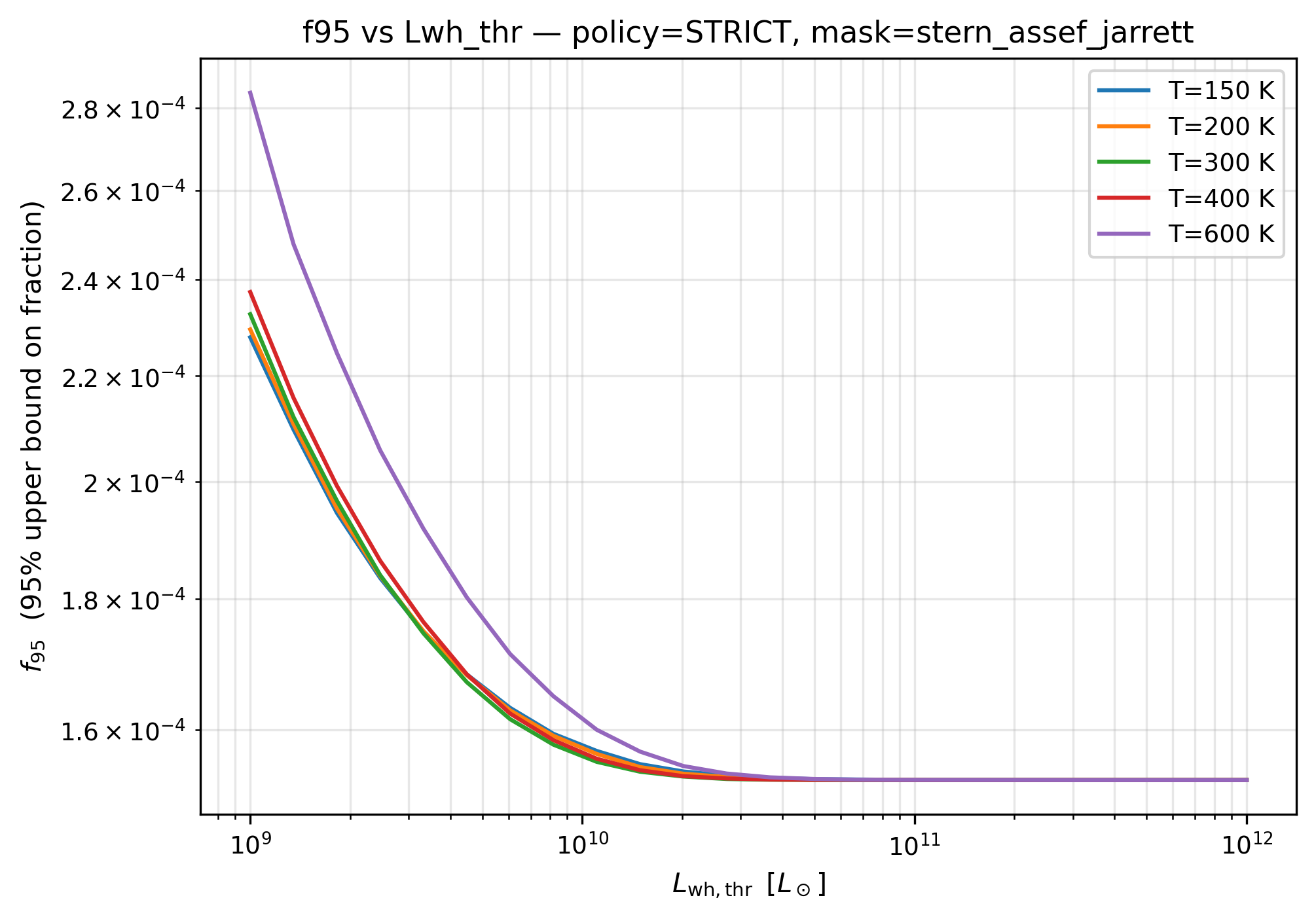}
  \caption{%
  95\% upper bounds on the fraction of galaxies, $f_{95}$, as a function of the waste heat luminosity threshold $L_{\rm wh,thr}$, for radiator temperatures $T=\{150,200,300,400,600\}$~K. The analysis uses the \textsc{Strict} sample (artifact--clean per band) with the combined Stern+Assef+Jarrett AGN/starburst masks applied. Upper limits are per object ``flux caps'' (W3/W4) with $k=3$ (i.e., $3\sigma$), and no stellar RJ subtraction unless otherwise noted. Curves are monotonic and converge where essentially all galaxies are sensitive; see Fig.~\ref{fig:neff} for the corresponding denominators.}
  \label{fig:f95-main}
\end{figure*}

\begin{figure*}[t!]
  \centering

  \includegraphics[width=0.6\textwidth]{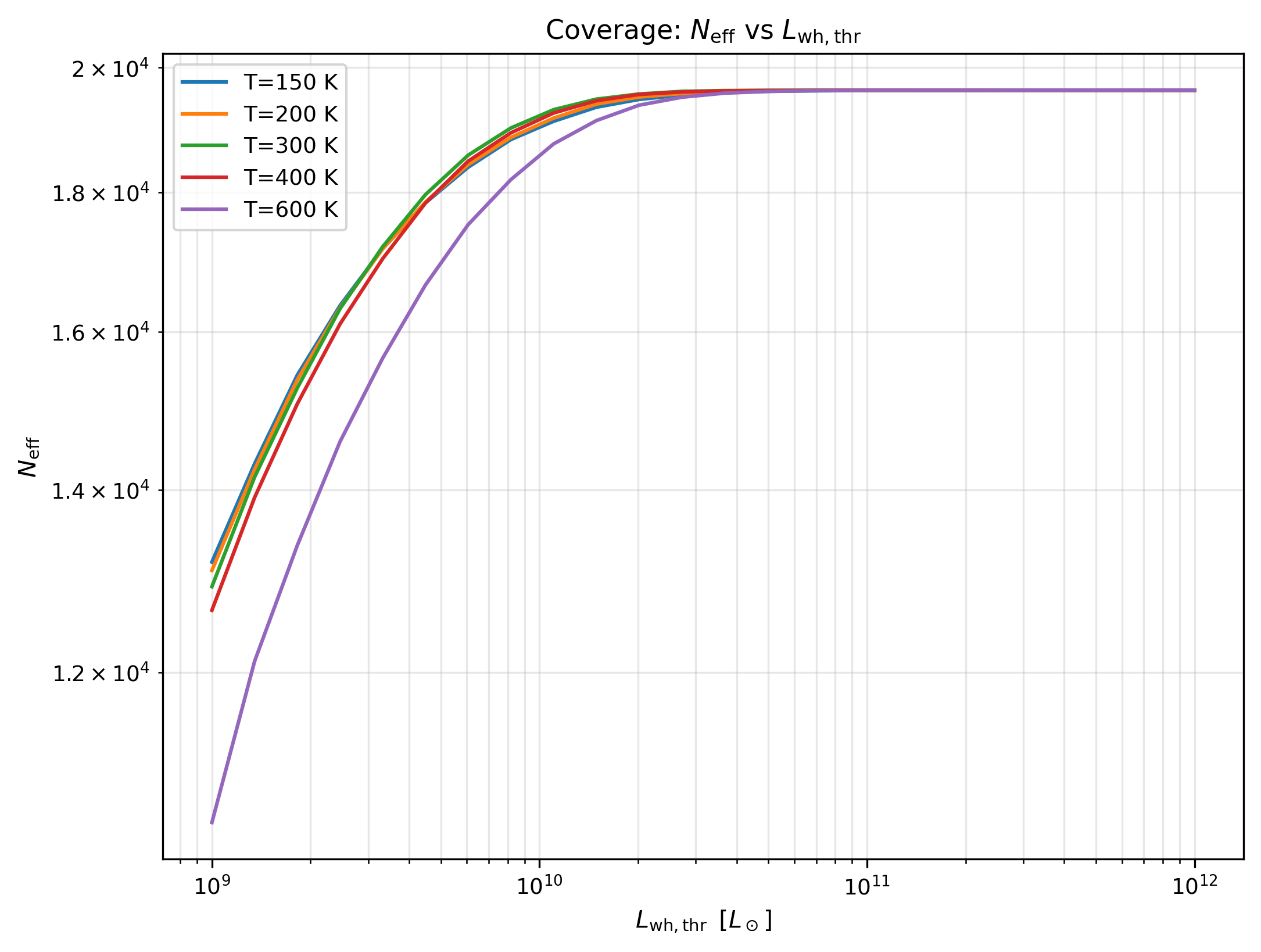}
  \caption{%
  Effective number of galaxies, $N_{\rm eff}$, that would yield a constraint at or below a given $L_{\rm wh,thr}$, for $T=\{150,200,300,400,600\}$~K (\textsc{Strict} + union mask). The rapid rise and saturation toward the parent sample size quantify the survey’s constraining power and directly underwrite Fig.~\ref{fig:f95-main} via $f_{95}\simeq 3/N_{\rm eff}$ in the zero event regime.}
  \label{fig:neff}
\end{figure*}

\begin{figure*}[t!]
  \centering

  \includegraphics[width=0.6\textwidth]{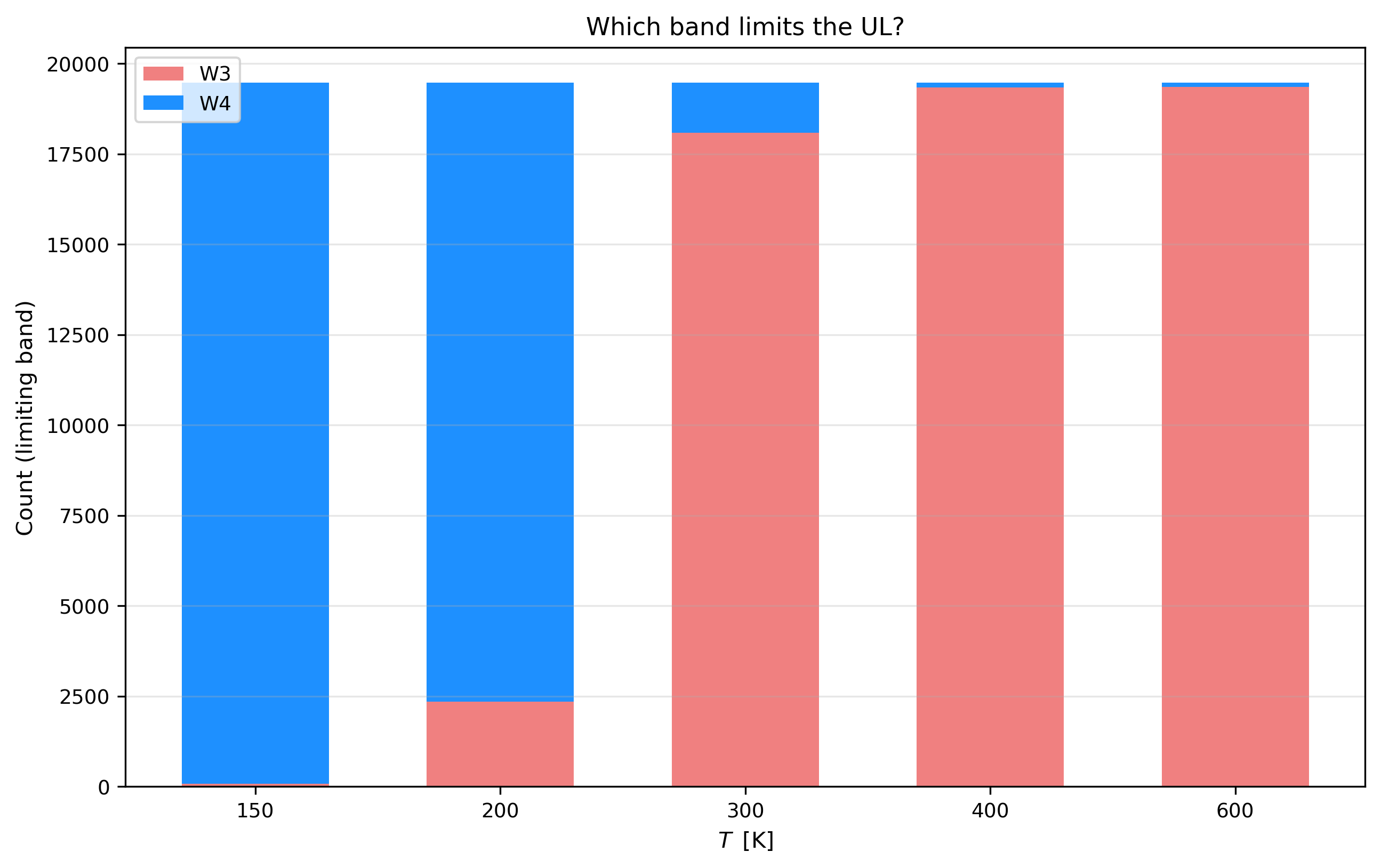}
 
  \caption{%
  Which band sets the tightest per object upper limit as a function of radiator temperature $T$. At $T\lesssim 200$~K, W4 (22\,$\mu$m) overwhelmingly limits; at $T\gtrsim 300$~K, W3 (12\,$\mu$m) dominates, consistent with Wien’s law and band sensitivities. These statistics explain the shapes of Fig.~\ref{fig:f95-main} and guide future observing strategy.}
  \label{fig:which-band}
\end{figure*}

For a fixed $T$ and threshold $L_{\rm wh,thr}$, we define the effective sample size
\begin{equation}
N_{\rm eff}(T,L_{\rm wh,thr}) \;\equiv\; \#\bigl\{\,\text{galaxies with } L_{\rm wh}^{\max}(T)\le L_{\rm wh,thr}\,\bigr\},
\end{equation}
i.e., the number of objects for which our data would have been sensitive to waste–heat luminosity at or above $L_{\rm wh,thr}$. With zero detections, the one–sided $95\%$ upper bound on the population fraction is well approximated by
\begin{equation}
f_{95}(T,L_{\rm wh,thr}) \;\simeq\; \frac{3}{N_{\rm eff}(T,L_{\rm wh,thr})},
\end{equation}
consistent with the Clopper–Pearson interval for the large $N_{\rm eff}$ in our sample \citep{ClopperPearson1934}. Figure~\ref{fig:f95-main} shows $f_{95}$ as a function of $L_{\rm wh,thr}$ for each $T$, while Fig.~\ref{fig:neff} shows the corresponding $N_{\rm eff}$ curves.

\begin{figure*}[t!]
  \centering
  \includegraphics[width=0.6\textwidth]{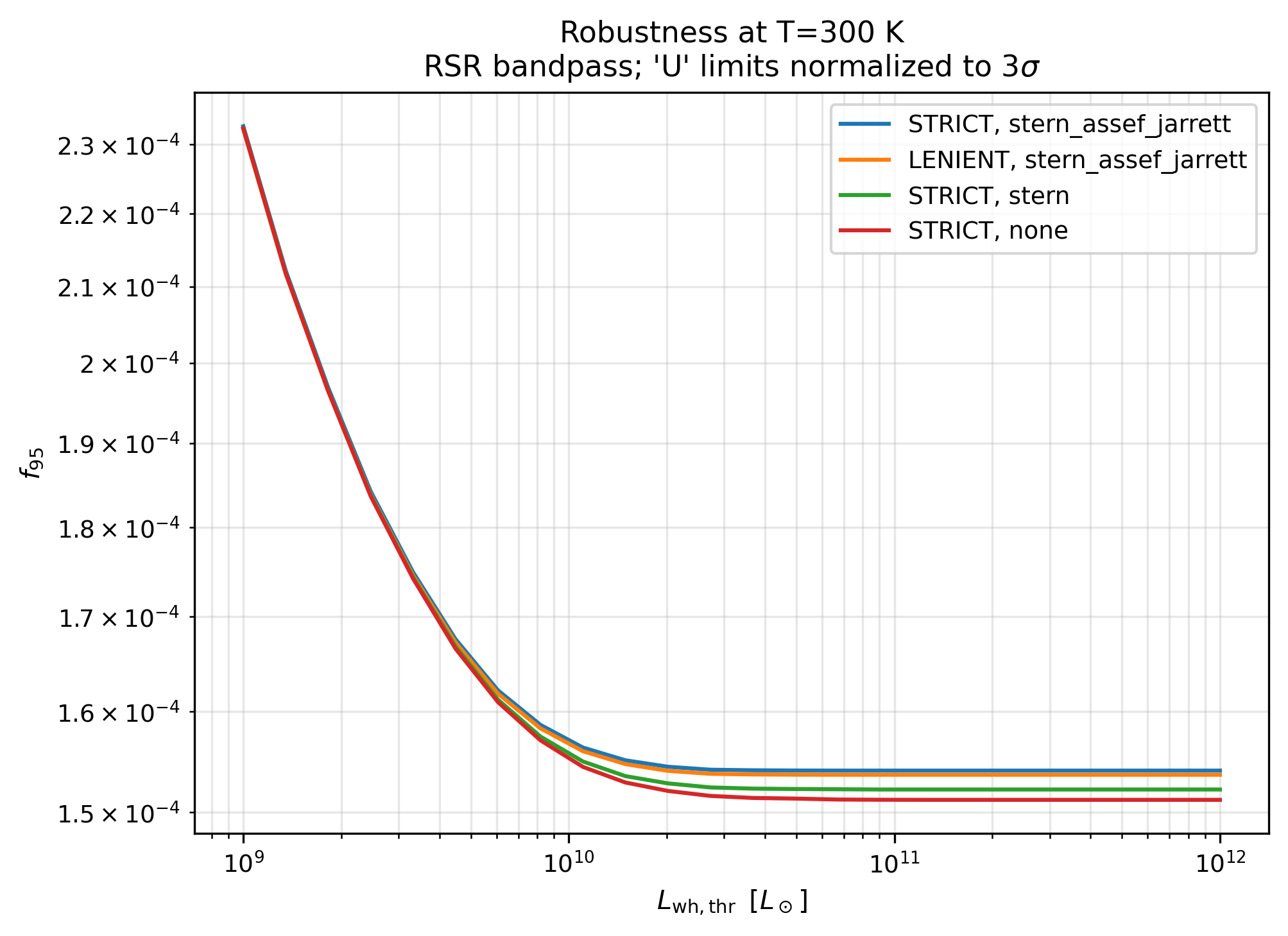}
  \caption{%
  Robustness of $f_{95}(L_{\rm wh,thr})$ at $T=300$~K to methodological choices. We compare \textsc{Strict} vs.\ \textsc{Lenient} (W4 relaxed) artifact policies and different masking strategies (none, Stern only, Stern+Assef, and the full Stern+Assef+Jarrett union). Curves are nearly indistinguishable across the threshold range, demonstrating that our main conclusions are insensitive to reasonable variations in artifact cuts and AGN/starburst masks.}
  \label{fig:f95-robust}
\end{figure*}

\begin{figure*}[t!]
  \centering
  \includegraphics[width=0.76\textwidth]{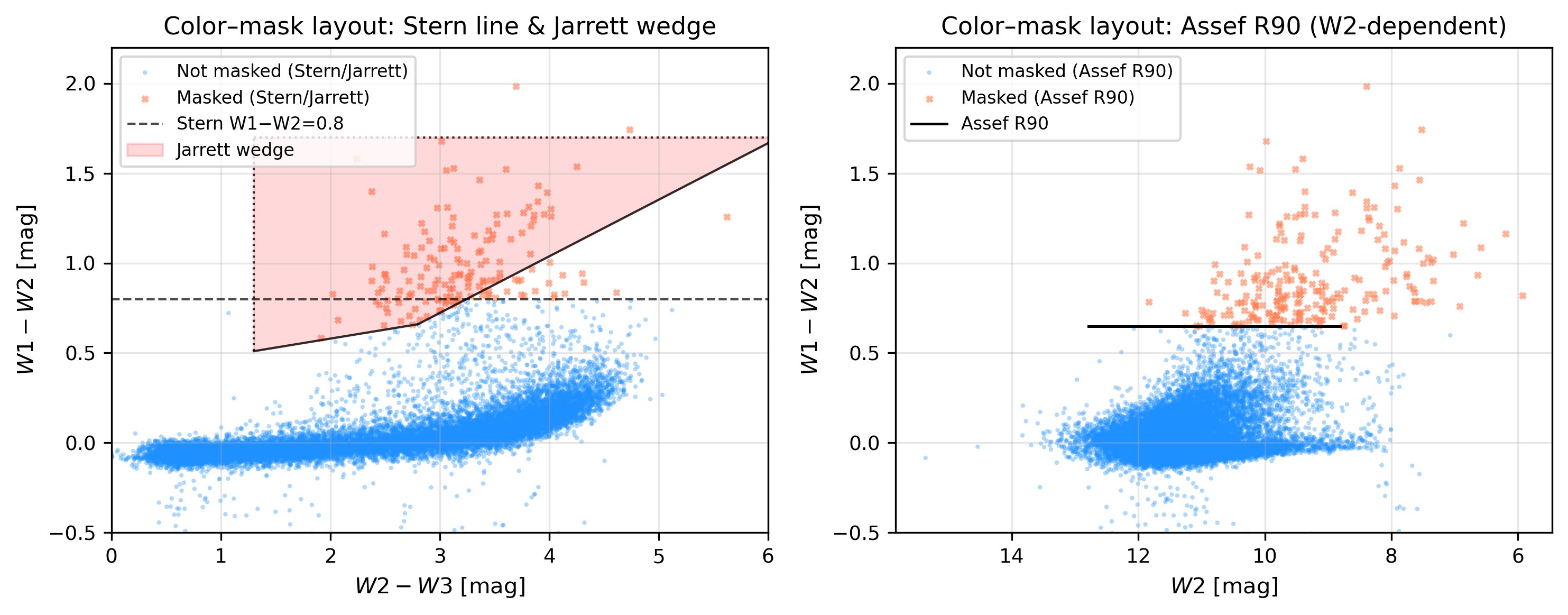}
\caption{%
Color–color diagrams and mask footprints. \textbf{Left:} $W1{-}W2$ versus $W2{-}W3$ with points colored by retained (blue) vs masked (orange) under the two masks. We overlay the \textbf{Stern} threshold $W1{-}W2=0.8$ (horizontal dashed) and the \textbf{Jarrett} wedge boundaries (solid), shading only the domain where the wedge applies ($W2{-}W3\ge 1.3$ and $W1{-}W2\le1.7$). \textbf{Right:} $W1{-}W2$ versus $W2$ with the \textbf{Assef R90} reliability cut, $(W1{-}W2)\ge \mathcal{R}_{90}(W2)$, which is not representable on the left-hand plane. The union mask is the logical OR of these three criteria.}

  \label{fig:color-mask}
\end{figure*}

\section{Results}\label{sec:results}

\subsection{Working Sample and Masks}\label{sec:results-sample}
Starting from the 2MRS parent list (\S\ref{sec:data}), cross–matching to \textit{CatWISE2020} and \textit{AllWISE} with the best–candidate protocol of \S\ref{sec:matching} yields a working set of $\sim2.2\times10^4$ galaxies with usable MIR photometry. The \textsc{Strict} sample requires clean per–band artifact flags for all four bands, whereas the \textsc{Lenient} variant retains W4 even if flagged (\S\ref{sec:policy}). We then apply the union of three independent MIR AGN/starburst masks (Stern, Assef R90, and a Jarrett–style wedge; \S\ref{sec:masks}). The retained sources occupy the locus of normal galaxies in WISE color–color space, while masked objects cluster in the expected AGN/starburst regions (Fig.~\ref{fig:color-mask}). After all selections, the number of galaxies with a valid per–object limit is
\emph{temperature–independent} and equals $N_{\rm valid}=19{,}617$ for the \textsc{Strict}+full–mask sample (Table~\ref{tab:quantiles}).

\begin{deluxetable*}{cccccc}
\tabletypesize{\footnotesize}
\tablecaption{Per–temperature distributions of per–galaxy waste–heat upper limits $L_{\rm wh}^{\max}$\label{tab:quantiles}}
\tablehead{
\colhead{$T$ [K]} & \colhead{$N_{\rm valid}$} &
\colhead{$p50$ [$L_\odot$]} & \colhead{$p90$ [$L_\odot$]} &
\colhead{$p95$ [$L_\odot$]} & \colhead{$p99$ [$L_\odot$]}
}
\startdata
150 & 19617 & 5.080e+08 & 4.090e+09 & 7.141e+09 & 1.769e+10 \\
200 & 19617 & 5.149e+08 & 4.106e+09 & 6.948e+09 & 1.597e+10 \\
300 & 19617 & 5.354e+08 & 3.938e+09 & 6.263e+09 & 1.334e+10 \\
400 & 19617 & 5.701e+08 & 4.184e+09 & 6.636e+09 & 1.433e+10 \\
600 & 19617 & 8.719e+08 & 6.421e+09 & 1.022e+10 & 2.220e+10 \\
\enddata
\tablecomments{%
For each radiator temperature $T$, we report the number of galaxies with a valid per–object limit ($N_{\rm valid}$) and the
median and high–quantile values ($p50,p90,p95,p99$) of $L_{\rm wh}^{\max}$ for the \textsc{Strict}+full–mask sample.
These per–object statistics set the scale for single–galaxy inferences and underwrite the population curves in Fig.~\ref{fig:f95-main}.
}
\end{deluxetable*}

\subsection{Per–object Upper–limit Distributions}\label{sec:results-single}
For each galaxy and each assumed radiator temperature $T\in\{150,200,300,400,600\}\,\mathrm{K}$, we compute a conservative upper limit $L_{\rm wh}^{\max}(T)$ using the W3/W4 ``flux-cap'' method of \S\ref{sec:pergal}, which includes bandpass integrated (RSR) color corrections and a $\sigma$--consistent handling of non detections. Table~\ref{tab:quantiles} summarizes the resulting distributions via the median and high quantiles. The distributions are tight and smooth across $T$, with the warmest case ($T=600$\,K) yielding the least constraining per object limits (as W3/W4 sample the Rayleigh Jeans tail), whereas $T=200$--$400$\,K form a narrow, more constraining family. These benchmarks enable direct comparison to prior WISE based technosignature searches.

\subsection{Population Upper Bounds: $f_{95}$ Curves}\label{sec:results-pop}
We aggregate per–galaxy limits into population constraints as a function of threshold, defining $N_{\rm eff}(T,L_{\rm wh,thr})$ and $f_{95}(T,L_{\rm wh,thr})\simeq 3/N_{\rm eff}$ (\S\ref{sec:pop}). Figure~\ref{fig:f95-main} shows the primary result: $f_{95}$ versus $L_{\rm wh,thr}$ for all five temperatures in the \textsc{Strict}+full–mask analysis. The curves are monotonic and converge at large $L_{\rm wh,thr}$, where essentially the entire sample is sensitive; the high–threshold plateau corresponds to
\[
f_{95}\;\approx\;\frac{3}{N_{\rm valid}}\;\!=\;\frac{3}{19{,}617}\;\simeq\;1.53\times 10^{-4},
\]
consistent with the Clopper–Pearson expectation at large $N_{\rm eff}$. At lower thresholds the curves separate in the expected sense: $T=600$\,K provides the weakest constraints, $T=150$–$200$\,K are limited by W4 depth, and $T=200$–$400$\,K give the strongest bounds.

Figure~\ref{fig:neff} presents the corresponding denominators, $N_{\rm eff}$ versus $L_{\rm wh,thr}$, which rise steeply and saturate near $N_{\rm valid}\approx1.96\times10^4$. Together, Figs.~\ref{fig:f95-main} and \ref{fig:neff} provide a complete summary of the population–level sensitivity of this search as a function of both temperature and threshold.

\subsection{Which Band Limits the Upper Limit?}\label{sec:results-band}
Because $L_{\rm wh}^{\max}(T)$ is the minimum of the W3– and W4–based caps (\S\ref{sec:pergal}), it is informative to record which band is limiting. Figure~\ref{fig:which-band} shows the counts of objects limited by W3 or by W4 as a function of $T$. At the coolest temperatures (150–200\,K), W4 (22\,$\mu$m) dominates the constraints, reflecting its proximity to the thermal peak and its leverage on cool dust–like SEDs. For $T\ge 300$\,K the balance flips: W3 (12\,$\mu$m) overwhelmingly sets the limit (the fraction W3–limited exceeds $\sim$90\%), consistent with W3 moving onto the spectral peak while W4 samples the Rayleigh–Jeans tail. This diagnostic explains the shapes of the $f_{95}$ curves and guides future observing strategies (e.g., deeper W4 for cool radiators; improved W3 calibration/depth for warm radiators).

\section{Robustness and Systematics}\label{sec:robust}

\subsection{Artifact Policy and Mask Choices}\label{sec:robust-policies}
Figure~\ref{fig:f95-robust} overlays $f_{95}(L_{\rm wh,thr})$ at $T=300$\,K for a policy/mask combinations. The \textsc{Lenient} policy (W4 relaxed) produces curves that are visually indistinguishable from \textsc{Strict}, as expected because (i) for $T\!\ge\!300$\,K the limiting band is overwhelmingly W3 (Fig.~\ref{fig:which-band}), and (ii) relaxing W4 only affects a minority of otherwise clean sources. Mask choices (none; Stern only; Stern+Assef; full union) likewise induce negligible changes: removing masks redistributes a small number of objects but does not alter the global sensitivity. Across all combinations, the asymptotic plateau remains fixed at $f_{95}\!\approx\!3/N_{\rm valid}\simeq1.53\times10^{-4}$ and the main conclusions of \S\ref{sec:results-pop} are unaffected. As an extreme bound, if all galaxies flagged by the AGN/starburst cuts were in fact KIII systems, our population constraints would revert to the “no‑mask’’ curves in Fig.~\ref{fig:f95-robust}; those are statistically indistinguishable from the baseline within our precision, so the limits would hardly change.

\subsection{Bandpass Color Corrections (RSR)}\label{sec:robust-rsr}
\begin{figure}
  \centering
  \includegraphics[width=0.6\linewidth]{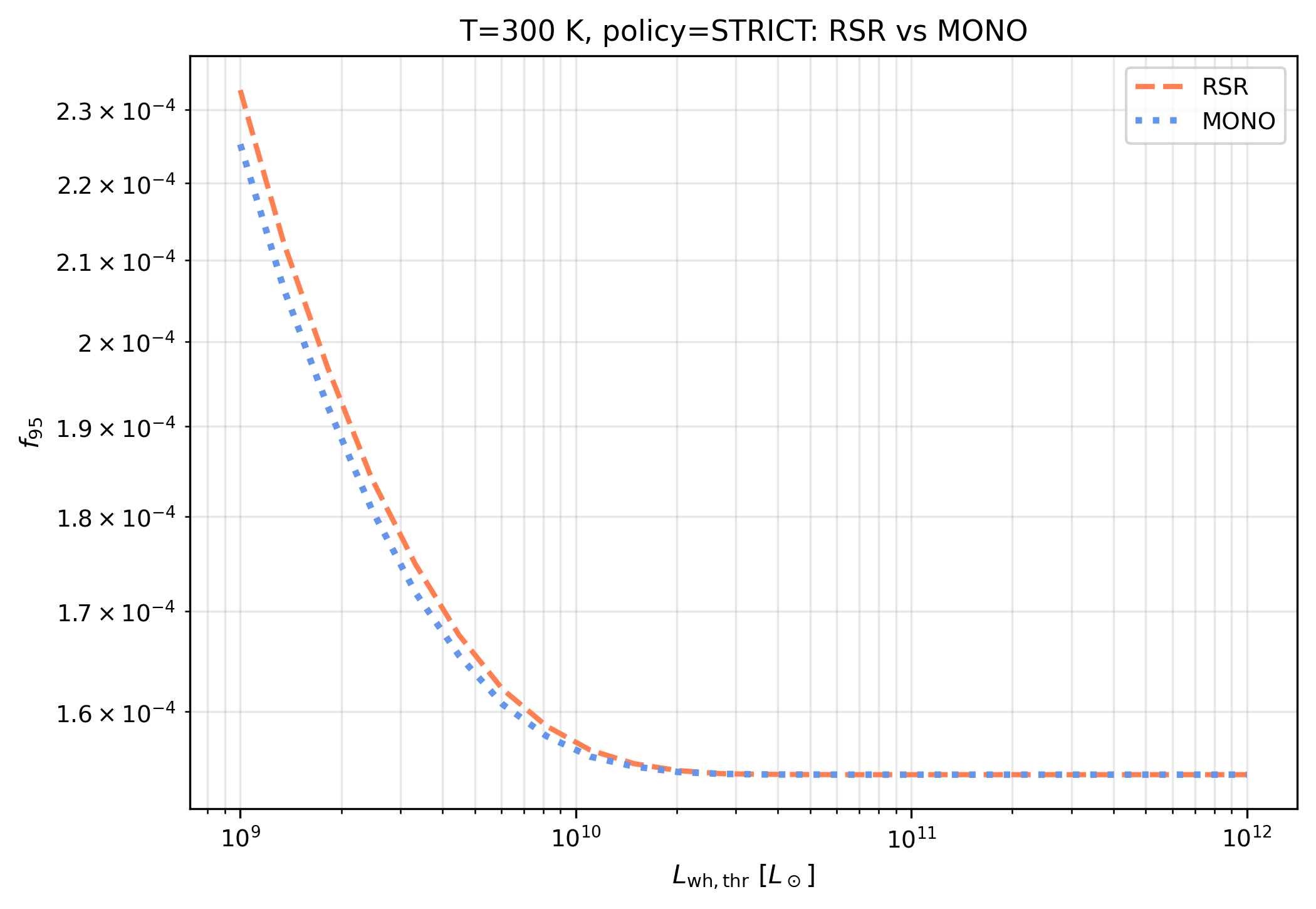}
  \caption{$f_{95}(L_{\rm wh,thr})$ at $T=300$\,K under RSR (bandpass–averaged) vs.\ MONO (monochromatic at the pivot).
  Curves differ at mid thresholds in the sense predicted by $\langle B_\nu\rangle/B_\nu(\nu_{\rm piv})$ and coincide at the high–threshold plateau set by $N_{\rm eff}$.}
  \label{fig:A1_rsr_mono_overlay}
\end{figure}

We test the robustness of our analysis to the treatment of the WISE passbands by comparing the bandpass-averaged RSR mapping to a monochromatic mapping at the pivot wavelength. Here “MONO’’ denotes a monochromatic mapping that replaces $\langle B_\nu(T)\rangle_b$ by $B_\nu(\nu_{\rm piv},T)$ at the band’s pivot frequency; “RSR’’ denotes the bandpass‑averaged mapping of Eq.~\ref{eq:bbavg}. Supplementary Fig.~\ref{fig:A1_rsr_mono_overlay} shows the resulting $f_{95}$ overlay at $T=300$\,K (Strict+Union mask). For the \textsc{Strict}+full–mask sample the median ratio $L_{\rm wh,RSR}^{\max}/L_{\rm wh,MONO}^{\max}$ is $\simeq$\,\textbf{+0.8\%} at $150$\,K, \textbf{+0.8\%} at $200$\,K (with a long high‑side tail to $\sim$+17\% among W3‑limited faint objects), \textbf{+16.6\%} at $300$\,K, \textbf{+9.1\%} at $400$\,K, and \textbf{$-1.7\%$} at $600$\,K. These shifts reflect the ratio $\langle B_\nu\rangle/B_\nu(\nu_{\rm piv})$ for a blackbody at the W3/W4 transition and move the \emph{mid‑threshold} portions of the $f_{95}$ curves accordingly. The \emph{high‑threshold} plateau, set by $N_{\rm eff}$, is unchanged.

\subsection[Non-detections and a Common k-sigma Convention]%
{Non-detections and a Common \texorpdfstring{$k\sigma$}{k-sigma} Convention}%
\label{sec:robust-ulimits}

Catalog non–detections (\texttt{ph\_qual}=U) are mapped to the same significance as detections via Eq.~\ref{eq:fluxcap}, with $\alpha_U\!\equiv\!k/k_U$ and $k\!=\!3$, $k_U\!\simeq\!2$. Varying $k_U$ over $[1.5,3]$ modifies only the faint‑end caps; the $f_{95}$ curves at bright thresholds and their convergence are insensitive because $f_{95}\!\propto\!1/N_{\rm eff}$ and $N_{\rm eff}$ is dominated by objects with firm detections or strong limits.

\subsection{Stellar RJ Subtraction}\label{sec:robust-rj}

We set $F^{\rm RJ}_{\nu,b}=0$ in Eq.~\ref{eq:fluxcap}, i.e., we do \emph{not} subtract any extrapolated stellar continuum at W3/W4. This choice is conservative, because any plausible stellar contribution would reduce $F^{\rm cap}_{\nu,b}$ and thus tighten the caps. As a robustness check we repeat the full analysis with a Rayleigh–Jeans extrapolation from W1/W2, parameterized as:
\[
F^{\rm RJ}_{\nu,b} \equiv A\,\nu_b^{2},\qquad 
A \equiv \tfrac12\!\left(\frac{F_{\nu,{\rm W1}}}{\nu_{\rm W1}^{2}}+\frac{F_{\nu,{\rm W2}}}{\nu_{\rm W2}^{2}}\right),
\]
so that \(F^{\rm RJ}_{\nu,b}\propto \nu_b^{2}\) (equivalently \(\propto \lambda_b^{-2}\)).
At $T{=}300$\,K the resulting $f_{95}(L_{\rm wh,thr})$ curve shifts downward by $\lesssim 2.2\%$ at mid thresholds and is unchanged at the high–threshold plateau; at $T{=}200$ and $400$\,K the shifts are $\lesssim 2.5\%$ and $\sim 0.6\%$, respectively.

\subsection{Extended–source Photometry and Systematic Floors}\label{sec:robust-extended}
AllWISE profile‑fit photometry can underestimate W3/W4 fluxes for large nearby galaxies. In an \emph{upper‑limit} analysis this bias makes $L_{\rm wh}^{\max}$ artificially \emph{small} (i.e., \emph{overly tight}, not conservative). Two features mitigate any residual impact: (i) the artifact policy removes poor fits and obvious artifacts; and (ii) for $T\!\ge\!300$\,K, W3 sets the limit for the vast majority of objects (Fig.~\ref{fig:which-band}), where extended‑source bias is less severe. We therefore expect any remaining bias to be subdominant relative to the bandpass and mask choices.

\begin{figure}[h!t]
  \centering
  \includegraphics[width=0.6\linewidth]{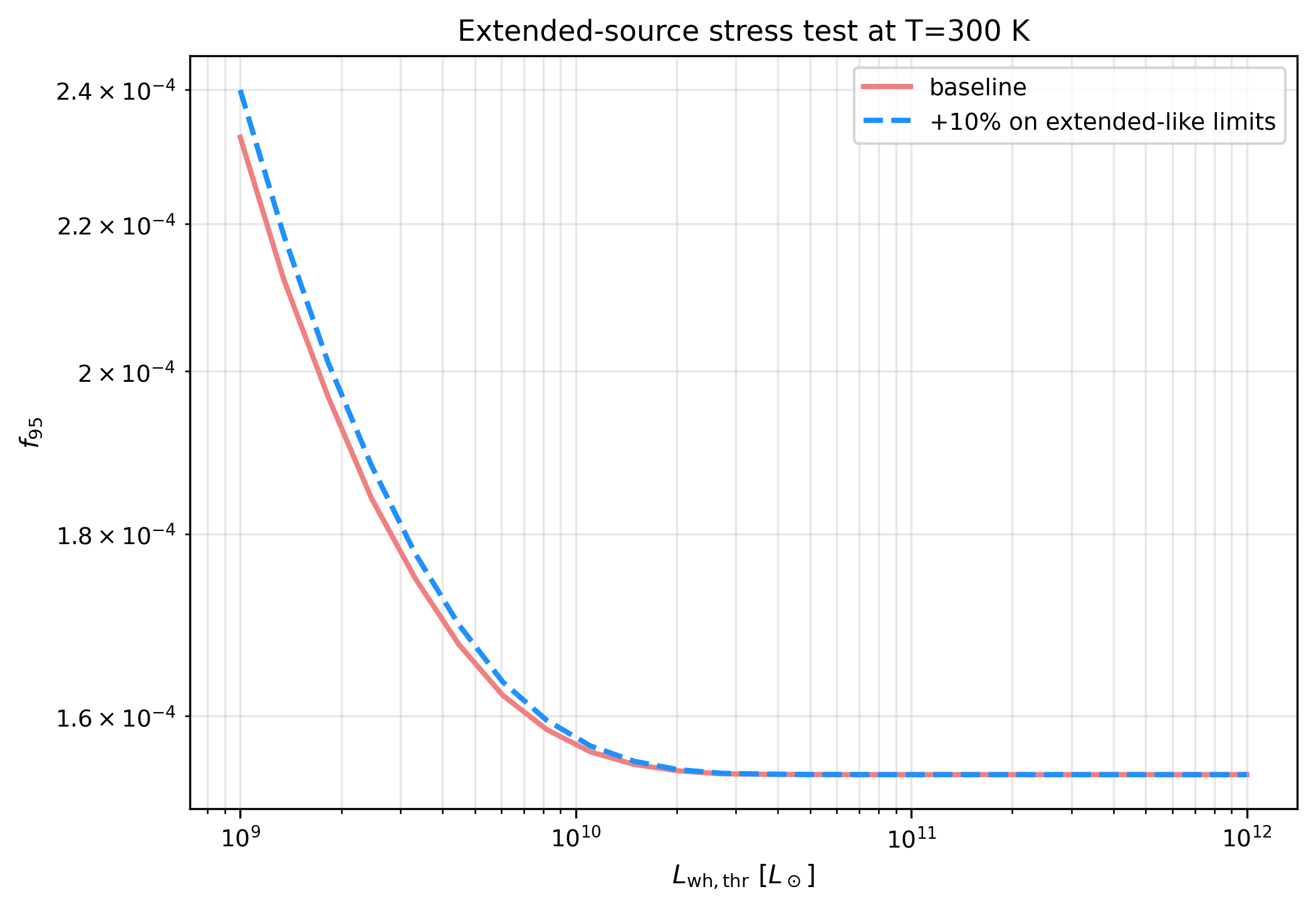}
  \caption{Extended‑source stress test at $T=300$\,K.
  We perturb the per–object $L_{\rm wh}^{\max}$ for “extended‑like’’ sources (large \(w1rchi2\), \(w2rchi2\), \(w3rchi2\), or \(w4rchi2\) by +10\% on the limiting band and recompute $f_{95}$. 
  The curve shifts by $\le 3\%$ at mid thresholds and leaves the high‑threshold plateau unchanged, confirming that any residual extended‑source bias is subdominant to the RSR bandpass effect and mask choices.}
  \label{fig:S2-ext-stress}
\end{figure}

A stress test perturbing W3/W4 by $+10\%$ for “extended‑like’’ objects (large \(w1rchi2\), \(w2rchi2\), \(w3rchi2\), or \(w4rchi2\), or \texttt{ext\_flg}$>$0) shifts the $T=300$\,K $f_{95}$ curve by $\le 3\%$ at mid thresholds and leaves the high‑threshold plateau unchanged (Fig.~\ref{fig:S2-ext-stress}).

\begin{figure*}[t]
  \centering
  \includegraphics[width=0.7\linewidth]{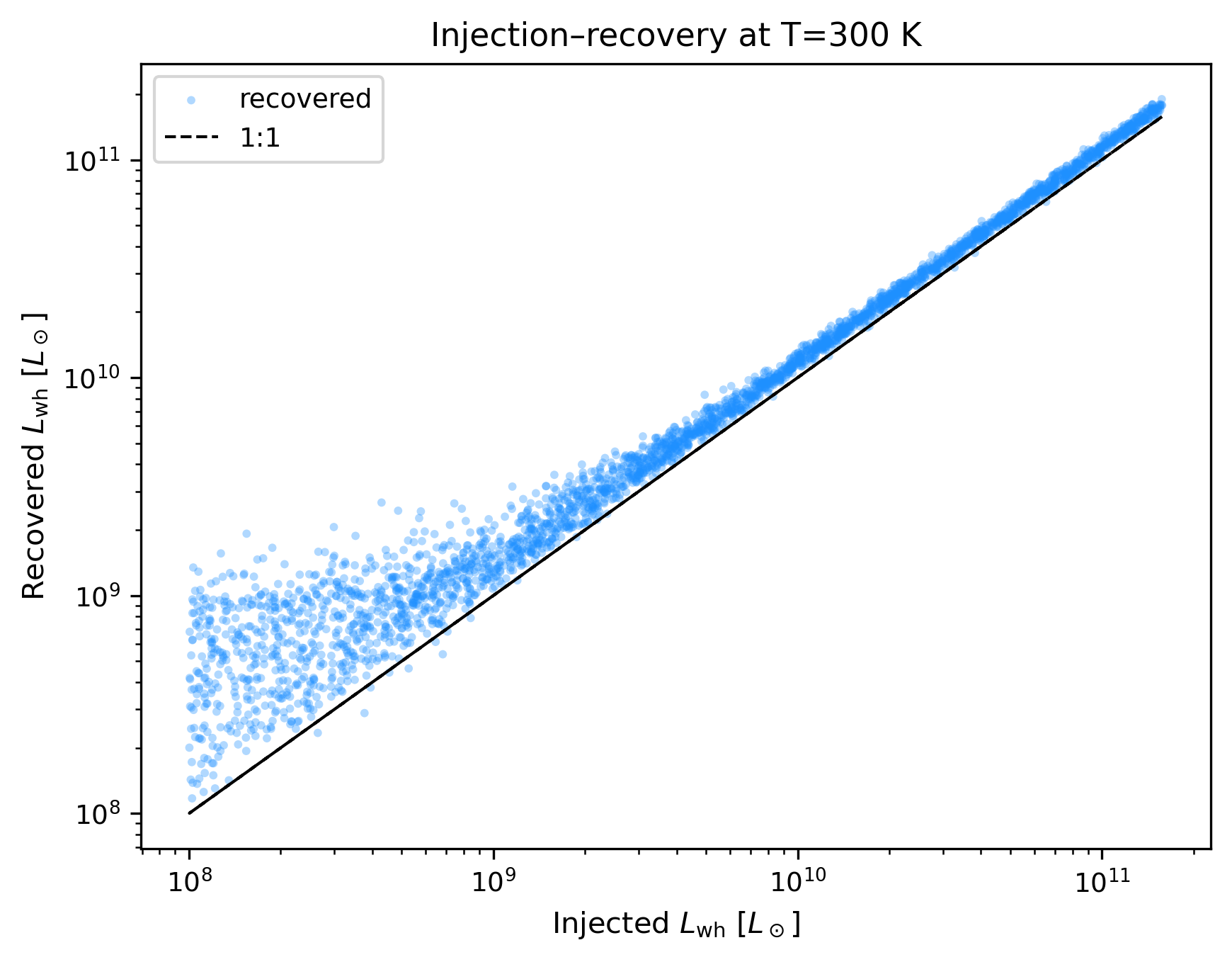}
  \caption{Injection recovery at $T=300$\,K.
  Each point is a synthetic galaxy: host MIR continuum (E/Sbc/SB family) plus an injected $T=300$\,K blackbody of bolometric luminosity $L_{\rm wh}$, integrated through the W3/W4 RSRs, with Gaussian noise and detection vs.\ non detection logic (cap at $k$, with threshold at $k_U$).
  The recovered value is obtained by mapping the per band caps to bolometric and taking the tighter of W3/W4.
  Points lie on the 1:1 line at high $L_{\rm wh}$ (unbiased recovery), while at low $L_{\rm wh}$ the recovered values are set by the cap floor and the un--subtracted host continuum, yielding conservative upper limits.
  }
  \label{fig:injrec-rigorous}
\end{figure*}

\subsection{Synthetic injection recovery validation}\label{sec:injrec}

We validate the end--to--end measurement model, including RSR bandpasses, noise, treatment of non--detections, and the $\min(\mathrm{W3},\mathrm{W4})$ limiting--band rule, with a Monte Carlo injection recovery test. Each trial draws a radiator temperature $T$, a luminosity distance $D_L$ sampled from the observed 2MRS distribution, and an injected waste--heat luminosity $L_{\rm wh}$. We then add a blackbody component of bolometric luminosity $L_{\rm wh}$ and temperature $T$ to a simple host mid--IR continuum.

The host continua (labeled E/Sbc/SB) are parametric toy SEDs defined in flux density $F_\nu$ at the WISE pivot frequencies to bracket dust poor through dust rich MIR host levels relevant for a W3/W4 based cap estimator. We write the host continuum as a Rayleigh Jeans stellar tail plus a featureless warm dust plateau,
\begin{equation}
F_{\nu,\rm host}(\nu_b) = A\left[\left(\frac{\nu_b}{\nu_{\rm W1}}\right)^2 + f_{\rm dust}\right],
\end{equation}
with $f_{\rm dust}=0$ (E; dust--poor), $f_{\rm dust}=0.3$ (Sbc; intermediate), and $f_{\rm dust}=1.0$ (SB; starburst--like). The normalization $A$ is chosen so that the resulting host W3/W4 fluxes lie in a realistic regime for the catalog uncertainties, ensuring that the detection/non--detection logic and the $k\sigma$ cap floor are exercised under conditions comparable to the data. These continua are intended as stress tests for the recovery pipeline, not as detailed MIR spectral models.

For each synthetic galaxy we integrate the combined host+blackbody spectrum through the W3 and W4 relative system responses (RSRs), add Gaussian noise comparable to the catalog errors, apply the same detection threshold at $k_U$ and the same $k\sigma$ cap construction as in the main analysis, and convert each band cap to a bolometric waste--heat limit. The recovered upper limit is then taken as $L_{\rm wh}^{\max}=\min(\mathrm{W3},\mathrm{W4})$, exactly matching the real--data workflow.

Figure~\ref{fig:injrec-rigorous} shows recovered versus injected $L_{\rm wh}$ at $T=300\,\mathrm{K}$. At high injected luminosity, the recovered values follow the 1:1 relation, demonstrating unbiased recovery when the waste--heat signal dominates the host continuum and noise. At low injected luminosity, the recovered values lie above the 1:1 line because non--negativity, the $k\sigma$ floor, and the host MIR continuum yield conservative caps, as intended. Across host families, the dust--poor E case yields the tightest recoveries, Sbc is modestly looser, and SB is loosest at low $L_{\rm wh}$ due to its stronger MIR continuum; however, such SB--like systems are preferentially removed by the union mask and therefore do not drive our reported population constraints.

\section{Discussion}\label{sec:discussion}

\subsection{What the Population Bounds Mean}\label{sec:disc-meaning}
Our primary result is a set of temperature–dependent population bounds, $f_{95}(T,L_{\rm wh,thr})$, that quantify what fraction of nearby galaxies could host waste–heat luminosity above a threshold $L_{\rm wh,thr}$ while remaining consistent with the observed WISE W3/W4 photometry under our $k\sigma$ flux–cap model. Because $f_{95}\!\propto\!1/N_{\rm eff}$ at bright thresholds, the high–threshold plateaus reflect only the effective sample size: with $N_{\rm valid}=19{,}617$ (Table~\ref{tab:quantiles}), all five curves converge to $f_{95}\simeq 3/N_{\rm valid}\simeq1.53\times10^{-4}$. At lower thresholds the curves separate in the physically expected order (Fig.~\ref{fig:f95-main}), encapsulating the different leverage of W3 and W4 on cool vs.\ warm blackbodies.

We intentionally refrain from imposing host–galaxy priors when presenting limits. For context, however, the single–object medians in Table~\ref{tab:quantiles} span $p50\simeq(5.0$–$8.6)\times10^{8}\,L_\odot$ from $T=150$–$600$\,K. If one adopts a representative stellar luminosity $L_\star\sim$ few $\times10^{10}\,L_\odot$, these caps correspond to order–percent fractional waste heat per galaxy; the exact mapping is population–dependent and not required for our conclusions.
\startlongtable
\begin{deluxetable*}{l c c c r c c c l}
\tabletypesize{\footnotesize}
\tablecaption{Top-50 galaxies with the loosest per object caps at $T=300$\,K (policy=\textsc{Strict}; masks=Stern+Assef~R90+Jarrett).}\label{tab:outliers}
\tablehead{
\colhead{ID} &
\colhead{RA [deg]} &
\colhead{Dec [deg]} &
\colhead{$z$} &
\colhead{$L_{\rm wh}^{\max}(300\,\mathrm{K})$ [$L_\odot$]} &
\colhead{Lim.\ band} &
\colhead{$W1{-}W2$ [mag]} &
\colhead{$W2{-}W3$ [mag]}
}
\startdata
16114086+5227270 & 242.92027 &  52.45754 & 0.029440367 & 48257844470 & W3 & 0.488 & 4.775  \\
00185089-1022364 &   4.71204 & -10.37680 & 0.027098747 & 39478049564 & W3 & 0.474 & 4.888  \\
18133982-5743312 & 273.41586 & -57.72527 & 0.017258606 & 36961814179 & W3 & 0.489 & 4.837  \\
01180836-4427432 &  19.53477 & -44.46194 & 0.022438857 & 36424901357 & W3 & 0.418 & 4.709  \\
02100957+3911253 &  32.53988 &  39.19037 & 0.017925734 & 36342849200 & W3 & 0.356 & 4.442  \\
23512674+2035105 & 357.86145 &  20.58624 & 0.018482787 & 34472687842 & W3 & 0.473 & 4.482  \\
16305653+0404583 & 247.73557 &   4.08289 & 0.024490276 & 32410231026 & W3 & 0.429 & 4.492  \\
00424586-2333418 &  10.69110 & -23.56160 & 0.022638995 & 32210083928 & W3 & 0.397 & 4.219  \\
22022283+1819072 & 330.59518 &  18.31872 & 0.027068726 & 29390958131 & W3 & 0.430 & 4.680  \\
09073082+1826057 & 136.87842 &  18.43487 & 0.029106803 & 27986125391 & W3 & 0.381 & 4.415  \\
14192669+7135177 & 214.86125 &  71.58823 & 0.025334193 & 26963601307 & W3 & 0.336 & 4.662  \\
\enddata
\tablecomments{%
Per object caps are computed at $T=300$\,K using the bandpass--averaged WISE RSR mapping and a $k{=}3$ cap convention (Eqs.~\ref{eq:fluxcap}--\ref{eq:lwh}).
The ``Lim.\ band'' column records whether W3 or W4 sets the bound. An ellipsis indicates a missing color (e.g., no W3 magnitude available).
\emph{The complete 50--row machine readable table is supplied as the ancillary file \texttt{outliers\_T300\_top50.csv}.}
A portion is shown here for guidance regarding its form and content.
}
\end{deluxetable*}
\subsection{Astrophysical Contaminants and the Transparency Table}\label{sec:disc-contaminants}
Our masking removes the most MIR–AGN–like and starburst–like colors (Stern, Assef R90, Jarrett; Fig.~\ref{fig:color-mask}), but the retained set still contains red systems with enhanced $W2{-}W3$ that are astrophysically mundane (e.g., dusty starbursts outside the strict wedge). The $T=300$\,K ``outliers'' (\textbf{top-50} loosest caps) are provided as a machine-readable ancillary table with this submission, \texttt{outliers\_T300\_top50.csv}. A key limitation of MIR-only Dyson-sphere searches is the degeneracy between warm waste heat and ordinary interstellar dust emission. Dust-rich, star-forming galaxies can exhibit strong W3/W4 continua that are spectrally similar to a few$\times10^2$\,K thermal component, inflating the flux caps and therefore yielding systematically looser $L_{\rm wh}^{\max}$ at $T\sim200$--$400$\,K even in the absence of any technosignature. This is the dominant astrophysical confounder behind the high-$L_{\rm wh}^{\max}$ tail and is why dust-poor (elliptical-like) systems provide markedly tighter constraints (\ref{sec:nulltests}).

\subsection{Systematics: What Matters and What Does Not}\label{sec:disc-systematics}
Three potential systematics were isolated and quantified in \S\ref{sec:robust}. First, the \emph{bandpass} treatment is material in the mid–threshold regime: using the WISE RSR color correction increases individual $L_{\rm wh}^{\max}$ by $\sim0$–1\% (150–200\,K), $\sim16\%$ (300\,K), $\sim9\%$ (400\,K), and $-1.7\%$ (600\,K) relative to a monochromatic mapping, as predicted by $\langle B_\nu\rangle/B_\nu(\nu_{\rm piv})$ and verified. Second, mapping catalog non–detections to the same $k\sigma$ as detections (Eq.~\ref{eq:fluxcap}) primarily affects the faint end and leaves $N_{\rm eff}$–dominated plateaus unchanged. Residual effects are subdominant because (i) the artifact policy removes poor photometry and (ii) W3 sets most limits for $T\!\ge\!300$\,K (Fig.~\ref{fig:which-band}). Additional checks—RJ subtraction, policy/mask variants—produce only modest, shape–preserving changes to the curves (Fig.~\ref{fig:f95-robust}).

\subsection{Dependence on Radiator Temperature and a Path to Improvement}\label{sec:disc-future}
The limiting–band statistics (Fig.~\ref{fig:which-band}) provide a practical roadmap. For \emph{cool} radiators ($T\!\le\!200$\,K), W4 (22\,$\mu$m) controls sensitivity; deeper W4 imaging, improved large–scale background modeling, or better artifact rejection would directly tighten the cool–$T$ curves. For \emph{warm} radiators ($T\!\ge\!300$\,K), W3 (12\,$\mu$m) dominates; improved W3 calibration and depth would yield the greatest leverage. Multi–wavelength anchoring of the long–wavelength SED (e.g., with far–IR photometry) would further reduce dependence on MIR color selection and sharpen the interpretation of red outliers.

\subsection{Empirical null tests across galaxy types}\label{sec:nulltests}
\begin{figure*}[h!t]
  \centering
  \includegraphics[width=0.6\linewidth]{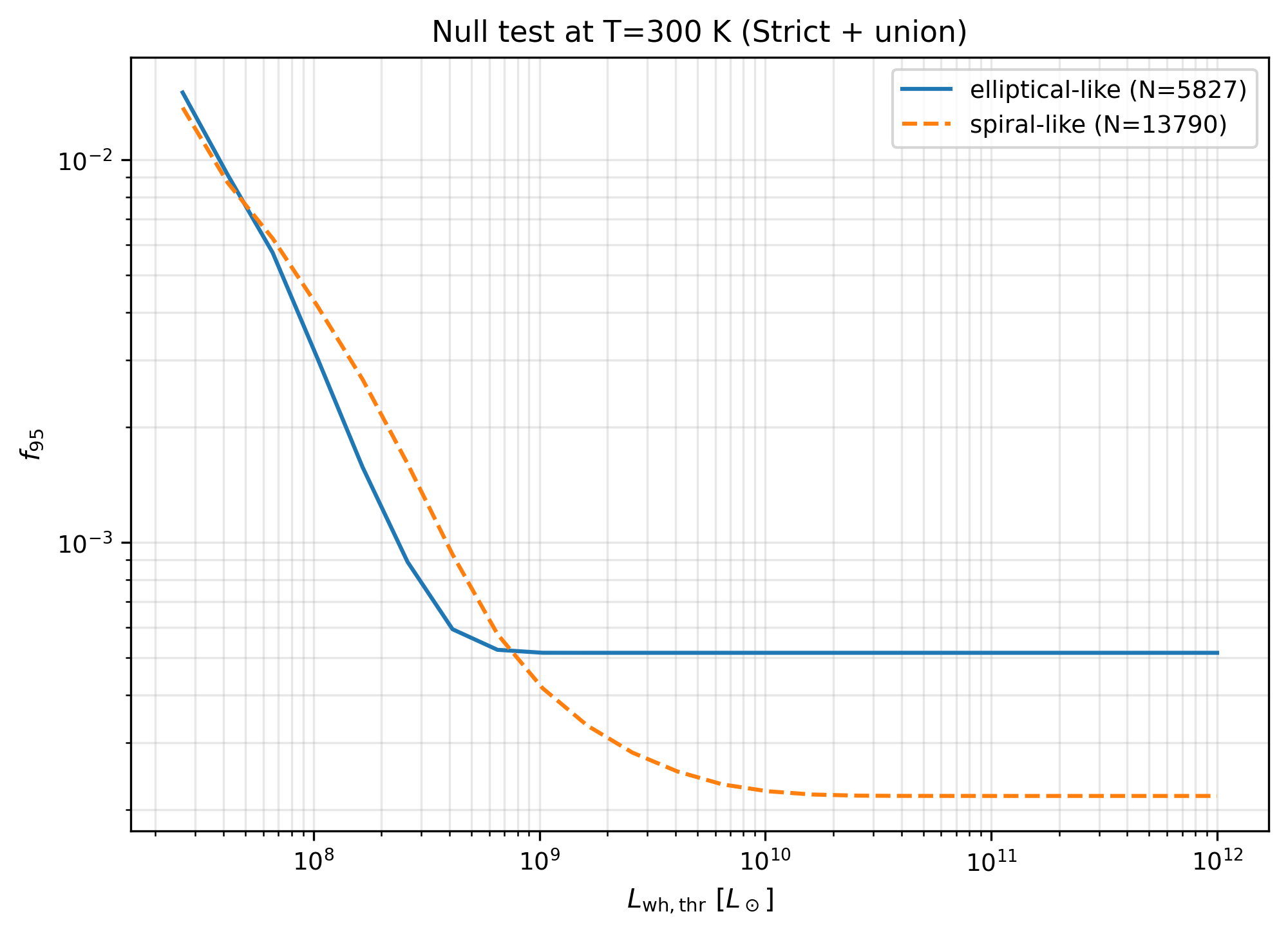}
  \caption{Empirical “null” test at $T=300$\,K (Strict + union masks).
  We split the sample by MIR color as a proxy for morphology (``elliptical-like'' with $W2{-}W3<1.0$ and ``spiral-like'' with $W2{-}W3\ge1.0$), excluding the Jarrett wedge. 
  The \emph{elliptical-like} subsample (\(N=5{,}827\)) sits on its binomial plateau across the plotted range with $f_{95}=3/N=5.15\times10^{-4}$, implying that all such galaxies have $L_{\rm wh}^{\max}\!\le10^{9}\,L_\odot$ at $T=300$\,K. 
  The \emph{spiral-like} subsample (\(N=13{,}790\)) behaves like the full sample, approaching the expected plateau $f_{95}=3/N=2.18\times10^{-4}$ at large thresholds. 
  This demonstrates that ordinary dust‑poor systems are intrinsically more constraining, and that our population bounds are not driven by a small set of red outliers.}
  \label{fig:S1-nulltests}
\end{figure*}

We split the \textsc{Strict}+union sample by MIR color as a proxy for morphology (``elliptical‑like’’ with $W2{-}W3<1.0$ and ``spiral‑like’’ with $W2{-}W3\ge1.0$), excluding the Jarrett wedge, and recompute $f_{95}$ at $T=300$\,K. All elliptical‑like galaxies ($N=5{,}827$) have $L_{\rm wh}^{\max}\!\le 10^{9}\,L_\odot$, so the curve sits on its plateau $3/N=5.15\times10^{-4}$ across the plotted range, whereas spiral‑like galaxies ($N=13{,}790$) behave like the full sample, approaching $3/N=2.18\times10^{-4}$ at large thresholds (Supplementary Fig.~\ref{fig:S1-nulltests}). This supports a “null’’ (no‑KIII) universe in which ordinary MIR color variations are driven by ISM rather than waste heat.

\subsection{Mapping to AGENT \texorpdfstring{$\alpha$}{alpha}}\label{sec:agent-alpha}
In the AGENT formalism \citep{Wright2014a}, the parameter $\alpha$ is the fraction of starlight reprocessed into waste heat, so $\alpha \approx L_{\rm wh}/L_\star$ for starlight-powered systems. Our per-object limits $L_{\rm wh}^{\max}$ therefore imply $\alpha^{\max} \approx L_{\rm wh}^{\max}/L_\star$. Because we did not model $L_\star$ per galaxy in this work, we present a fiducial translation using $L_\star = 3\times10^{10}\,\LsunNom$ (Milky Way-like): the median caps in Table~\ref{tab:quantiles} correspond to $\alpha^{\max}_{\rm median} \approx 1.7\%$ (150--300\,K), $1.9\%$ (400\,K), and $2.9\%$ (600\,K). A future extension using, e.g., 2MRS $K_s$ photometry to estimate $L_\star$ per galaxy would yield a distribution of $\alpha^{\max}$ values; none of our population conclusions rely on this additional modeling.

\section{Conclusions}\label{sec:conclusions}
We presented a WISE based search for Dysonian waste heat in nearby galaxies. The analysis (i) implements bandpass integrated relative spectral response (RSR) color corrections for W3/W4, (ii) treats non detections with a $\sigma$--consistent flux cap, and (iii) prefers profile--fit photometry, with extended source photometry assessed as a systematic. Applied to a $\sim 2.2\times 10^{4}$--object working sample, the \textsc{Strict}+full--mask subset yields $N_{\mathrm{valid}}=19{,}617$ per object limits across $T\in\{150,200,300,400,600\}\,\mathrm{K}$.

Our main findings are:
\begin{enumerate}
\item The population bounds $f_{95}(T,L_{\rm wh,thr})$ are monotonic in threshold and converge at large $L_{\rm wh,thr}$ to the binomial plateau $f_{95}\!\approx\!3/N_{\rm valid}\simeq1.53\times10^{-4}$ (Figs.~\ref{fig:f95-main}, \ref{fig:neff}).
\item The limiting band transitions from W4 at $T\!\le\!200$\,K to W3 at $T\!\ge\!300$\,K (Fig.~\ref{fig:which-band}), explaining the families of curves and highlighting where deeper data would most improve sensitivity.
\item Proper W3/W4 bandpass treatment matters at mid thresholds: relative to a monochromatic mapping, the RSR correction raises individual caps by $\sim$16\% at 300\,K (and $\sim$9\% at 400\,K), while leaving the high–threshold plateau fixed by $N_{\rm eff}$.
\item Mask choices, artifact policy (STRICT vs.\ LENIENT), and RJ subtraction induce only modest, shape–preserving changes; the conclusions are robust (Fig.~\ref{fig:f95-robust}).
\end{enumerate}
At the population level, the one-sided 95\% upper limit reaches $\simeq 1/6500$ (i.e., $f_{95}\!\approx\!1.53\times 10^{-4}$) at high thresholds, set purely by sample size. Interpreted via AGENT, typical \emph{per-galaxy} caps are $\alpha \lesssim (1.7$--$2.9)\%$ for $T{=}150$--$600$\,K when scaled to a Milky Way-like $L_\star$.

\section{Supplementary}

\begin{figure}[t]
  \centering
  \includegraphics[width=0.6\linewidth]{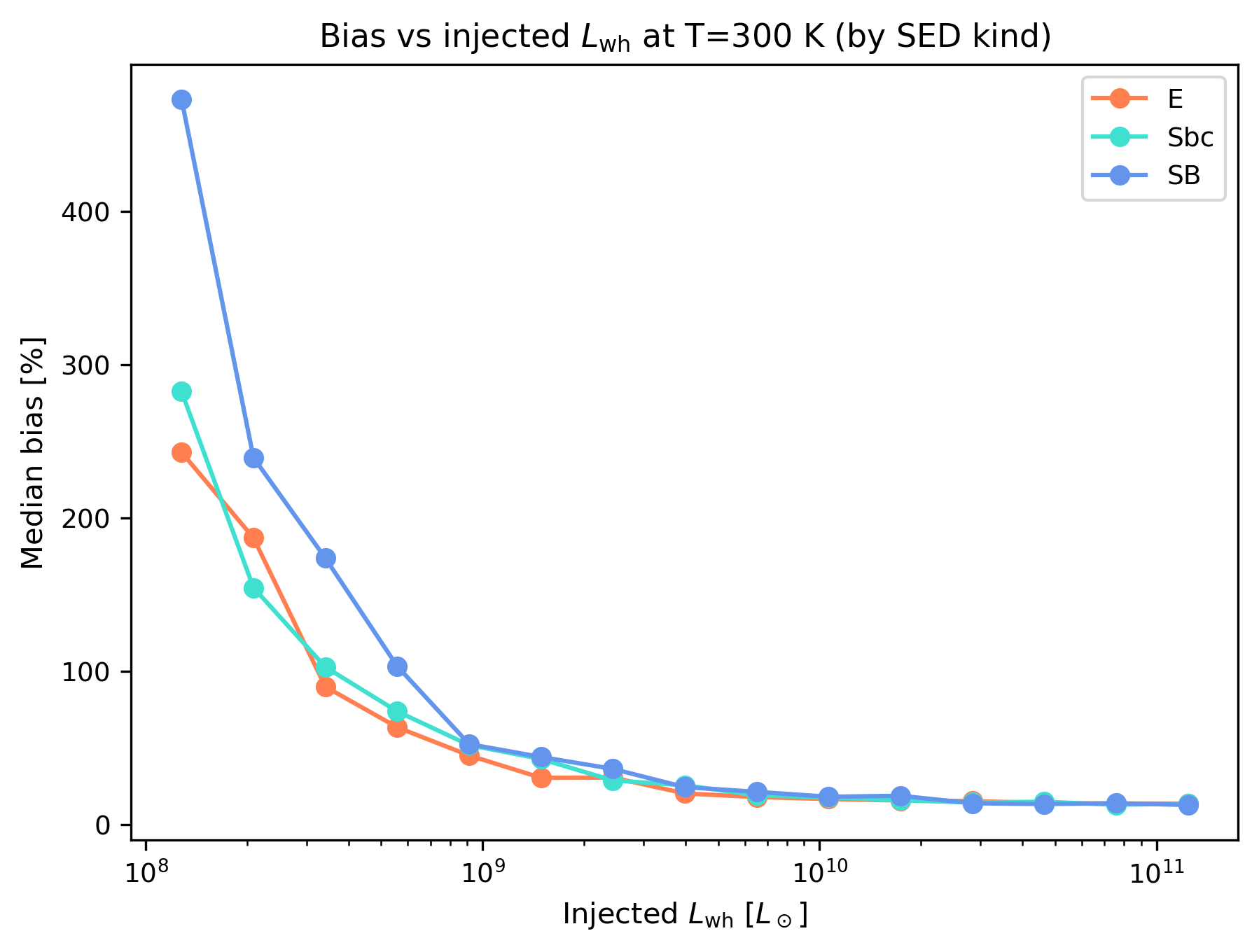}
  \caption{
  Median bias versus injected $L_{\rm wh}$ at $T=300$\,K, split by host SED family.
  Positive bias at low $L_{\rm wh}$ is the expected behavior of a cap estimator in the presence of non detections (floor at $\approx k\sigma$) and un--subtracted host MIR continuum.
  The three SED families converge toward small bias at high $L_{\rm wh}$. As expected, the bias is largest for SB at the faint end (strongest MIR continuum) and decreases toward small values at high $L_{\rm wh}$, where the cap floor is negligible relative to the injected signal.
  }
  \label{fig:injrec-bias}
\end{figure}

\begin{figure}
  \centering
  \includegraphics[width=0.6\linewidth]{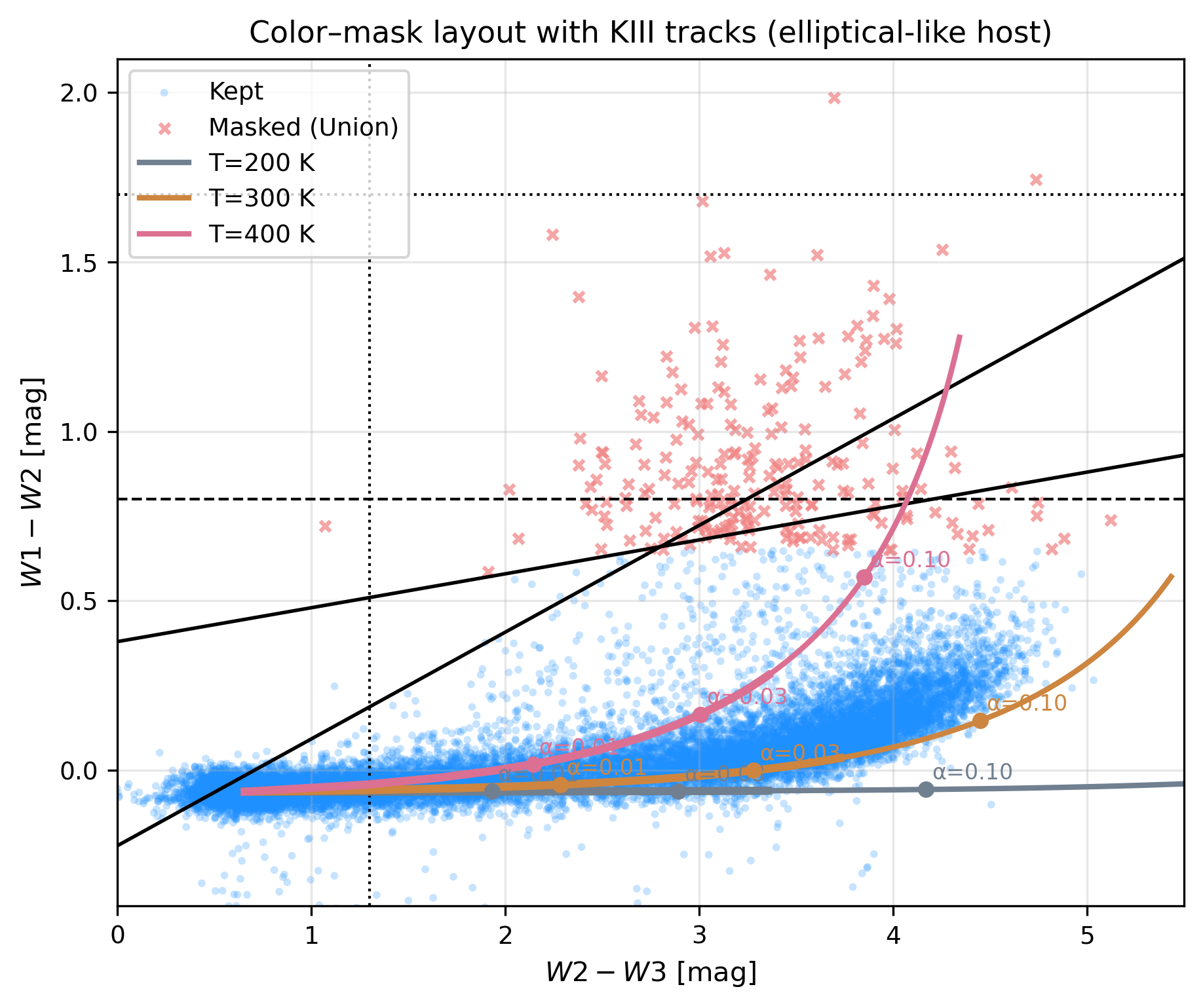}
  \caption{WISE $W1{-}W2$ versus $W2{-}W3$ (Vega) color--color diagram for the 2MRS cross matched galaxy sample, shown to visualize how our MIR AGN/starburst rejection interacts with plausible galaxy scale waste heat (KIII) spectral energy distributions. Blue circles are galaxies \emph{retained} for the analysis, while orange crosses are galaxies \emph{masked} by the adopted MIR AGN/starburst criteria (union of the Stern, Assef R90, and Jarrett diagnostics; see \S\ref{sec:masks}). The black guide lines show the commonly used MIR selection boundaries: the Stern threshold $W1{-}W2=0.8$ (horizontal dashed), and the Jarrett ``wedge'' boundaries (solid oblique lines) together with its additional limits $W2{-}W3\ge 1.3$ and $W1{-}W2\le 1.7$ (vertical and horizontal dotted lines, respectively). Overplotted colored curves are illustrative KIII ``tracks'' for an \emph{elliptical--like (dust poor) host} whose stellar MIR continuum is progressively supplemented by a single temperature waste heat blackbody at $T=\{200,300,400\}\,$K. The parameter $\alpha$ denotes the fractional waste heat luminosity in the AGENT sense (i.e., the fraction of the host’s bolometric power reprocessed and re--radiated thermally as waste heat at temperature $T$), and labeled points mark representative values along each track. The tracks are \emph{curved} (rather than straight vectors) because increasing $\alpha$ does not only raise the long wavelength emission (driving redder $W2{-}W3$), but at sufficiently large $\alpha$ it also alters the short wavelength bands: the stellar contribution to $W1$ and $W2$ is diluted/attenuated while the waste heat component can contribute non--negligibly in $W2$ for warmer radiators, changing $W1{-}W2$ and producing a non--linear trajectory in color space. This figure therefore provides a direct visual check of whether putative KIII--like SEDs would lie inside the AGN/starburst loci targeted by the union mask, and clarifies the interpretation of our robustness tests with and without these masks.}
  \label{fig:S4-color-arrow}
\end{figure}

\clearpage

\section*{Acknowledgments}
We thank the reviewer for their thoughtful and constructive reviews. 
This work was supported by the National Key R\&D Program of China (No.\ 2024YFA1611804), 
the China Manned Space Program (CMS-CSST2025-A01), the National SKA Program of China under Grant No. 2025SKA0120104, the Shandong Provincial Natural Science Foundation (ZR2024QA180), 
and the Scientific Research Fund of Dezhou University (4022504019). 
This research has made use of the NASA/IPAC Infrared Science Archive (IRSA), 
which is funded by the National Aeronautics and Space Administration and operated by 
the California Institute of Technology. 
This publication makes use of data products from the \textit{Wide-field Infrared Survey Explorer} and \textit{NEOWISE}. 
WISE and NEOWISE are funded by NASA and are a joint project of the University of California, Los Angeles, 
and the Jet Propulsion Laboratory/California Institute of Technology.

\section*{Data Availability}
A single machine-readable ancillary file, \texttt{outliers\_T300\_top50.csv} (top-50 loosest caps at $T=300$\,K), accompanies this article. AllWISE Source Catalog data are available via IRSA (DOI: \doi{10.26131/IRSA1}). CatWISE2020 catalog data are available via IRSA (DOI: \doi{10.26131/IRSA551}).

\facilities{IRSA, WISE}

\appendix
\section{Workflow pseudo-code}\label{app:workflow}
\begin{verbatim}
INPUT: 2MRS positions/redshifts, AllWISE Source Catalog, CatWISE2020
FOR each 2MRS galaxy:
  - Cone-search AllWISE and CatWISE2020 within 5"
  - Select one "best" counterpart per catalog by:
      (1) cleanest artifact flags; (2) smallest separation; (3) W1 profile-fit quality
  - Adopt W1/W2 (baseline): prefer CatWISE if clean, else AllWISE if clean
  - Adopt W3/W4 (constraints): AllWISE (retain W4 based on policy)
Apply artifact policy:
  STRICT: W1 TO W4 ALL clean
  LENIENT: W1 TO W3 clean AND W4 kept even if flagged (marked)
Compute colors; apply AGN/starburst masks:
  - Stern: (W1-W2) >= 0.8 if W2 < 15.05
  - Assef R90: (W1-W2) >= R90(W2)
  - Jarrett wedge: domain + two oblique boundaries
Select retained sample (union mask if desired)
FOR each T in {150,200,300,400,600} K:
  FOR each constraining band b in {W3,W4}:
    - Form 3-sigma cap F_nu,b^cap using detections/non-detections consistently
    - Convert to bolometric cap with WISE RSR: F_bol^(b)(T)
    - L_wh^(b)(T) = 4*pi D_L^2 F_bol^(b)(T)
  - L_wh^max(T) = min[L_wh^(W3)(T), L_wh^(W4)(T)]
Aggregate:
  - Distributions of L_wh^max(T) (p50, p90, ...)
  - N_eff(T, L_thr) = count[L_wh^max(T) <= L_thr]
  - f_95(T, L_thr) ~= 3 / N_eff(T, L_thr)  (zero-event upper bound)
ROBUSTNESS:
  - Policy variants (STRICT/LENIENT), masks (none/Stern/Stern+Assef/union)
  - RSR vs monochromatic at pivot, RJ subtraction ON/OFF
  - Extended-source stress (+10% on limiting band)
\end{verbatim}

\bibliography{sample701}{}

@article{Dyson1960,
  author  = {Dyson, Freeman J.},
  title   = {Search for Artificial Stellar Sources of Infrared Radiation},
  journal = {Science},
  year    = {1960},
  volume  = {131},
  number  = {3414},
  pages   = {1667--1668},
  doi     = {10.1126/science.131.3414.1667}
}

@article{Wright2010,
  author  = {Wright, E. L. and Eisenhardt, P. R. M. and Mainzer, A. K. and Ressler, M. E. and Cutri, R. M. and Jarrett, T. and {et~al.}},
  title   = {The Wide-field Infrared Survey Explorer (WISE): Mission Description and Initial On-orbit Performance},
  journal = {AJ},
  year    = {2010},
  volume  = {140},
  number  = {6},
  pages   = {1868--1881},
  doi     = {10.1088/0004-6256/140/6/1868}
}

@article{Mainzer2014,
  author  = {Mainzer, A. and Bauer, J. and Cutri, R. M. and Grav, T. and Masiero, J. and Beck, R. and {et~al.}},
  title   = {Initial Performance of the NEOWISE Reactivation Mission},
  journal = {ApJ},
  year    = {2014},
  volume  = {792},
  number  = {1},
  pages   = {30},
  doi     = {10.1088/0004-637X/792/1/30}
}

@article{Huchra2012,
  author  = {Huchra, J. P. and Macri, L. M. and Masters, K. L. and {et~al.}},
  title   = {The 2MASS Redshift Survey—Description and Data Release},
  journal = {ApJS},
  year    = {2012},
  volume  = {199},
  pages   = {26},
  doi     = {10.1088/0067-0049/199/2/26}
}

@article{Marocco2021,
  author  = {Marocco, F. and Eisenhardt, P. R. M. and Fowler, J. W. and {et~al.}},
  title   = {The CatWISE2020 Catalog},
  journal = {ApJS},
  year    = {2021},
  volume  = {253},
  pages   = {8},
  doi     = {10.3847/1538-4365/abd805}
}

@misc{Cutri2013,
  author       = {Cutri, R. M. and Wright, E. L. and Conrow, T. and Fowler, J. and {et~al.}},
  title        = {Explanatory Supplement to the AllWISE Data Release Products},
  howpublished = {IPAC/Caltech},
  year         = {2013}
}

@misc{AllWISEExplanatory,
  author       = {Cutri, R. M. and {AllWISE Team}},
  title        = {AllWISE Data Release Explanatory Supplement (2014 Update)},
  howpublished = {IPAC/Caltech},
  year         = {2014}
}

@article{Stern2012,
  author  = {Stern, D. and Assef, R. J. and Benford, D. J. and {et~al.}},
  title   = {Mid-infrared Selection of Active Galactic Nuclei with the Wide-field Infrared Survey Explorer. I. Characterizing WISE-selected Active Galactic Nuclei in COSMOS},
  journal = {ApJ},
  year    = {2012},
  volume  = {753},
  number  = {1},
  pages   = {30},
  doi     = {10.1088/0004-637X/753/1/30}
}

@article{Assef2018,
  author  = {Assef, R. J. and Stern, D. and Noirot, G. and Jun, H. D. and Cutri, R. M. and Eisenhardt, P. R. M.},
  title   = {The WISE AGN Catalog},
  journal = {ApJS},
  year    = {2018},
  volume  = {234},
  number  = {2},
  pages   = {23},
  doi     = {10.3847/1538-4365/aaa00a}
}

@article{Jarrett2011,
  author  = {Jarrett, T. H. and Cohen, M. and Masci, F. and {et~al.}},
  title   = {The Spitzer–WISE Survey of the Ecliptic Poles},
  journal = {ApJ},
  year    = {2011},
  volume  = {735},
  number  = {2},
  pages   = {112},
  doi     = {10.1088/0004-637X/735/2/112}
}

@article{Wright2014a,
  author  = {Wright, J. T. and Mullan, B. and Sigurdsson, S. and Povich, M. S.},
  title   = {The {\^{G}} Infrared Search for Extraterrestrial Civilizations with Large Energy Supplies. I. Background and Justification},
  journal = {ApJ},
  year    = {2014},
  volume  = {792},
  number  = {1},
  pages   = {26},
  doi     = {10.1088/0004-637X/792/1/26}
}

@article{Wright2014b,
  author  = {Wright, J. T. and Griffith, R. L. and Sigurdsson, S. and Povich, M. S. and Mullan, B.},
  title   = {The {\^{G}} Infrared Search for Extraterrestrial Civilizations with Large Energy Supplies. II. Framework, Strategy, and First Results},
  journal = {ApJ},
  year    = {2014},
  volume  = {792},
  number  = {1},
  pages   = {27},
  doi     = {10.1088/0004-637X/792/1/27}
}

@article{Griffith2015,
  author  = {Griffith, R. L. and Wright, J. T. and Maldonado, J. and Povich, M. S. and Sigurdsson, S. and Mullan, B.},
  title   = {The {\^{G}} Infrared Search for Extraterrestrial Civilizations with Large Energy Supplies. III. The Reddest Extended Sources in WISE},
  journal = {ApJS},
  year    = {2015},
  volume  = {217},
  number  = {2},
  pages   = {25},
  doi     = {10.1088/0067-0049/217/2/25}
}

@article{Lang2014,
  author  = {Lang, Dustin},
  title   = {unWISE: unblurred coadds of the WISE imaging},
  journal = {AJ},
  year    = {2014},
  volume  = {147},
  number  = {5},
  pages   = {108},
  doi     = {10.1088/0004-6256/147/5/108}
}

@article{Meisner2017,
  author  = {Meisner, A. M. and Lang, D. and Schlegel, D. J.},
  title   = {Full-depth Coadds of the WISE and NEOWISE Imaging},
  journal = {AJ},
  year    = {2017},
  volume  = {153},
  number  = {2},
  pages   = {38},
  doi     = {10.3847/1538-3881/153/2/38}
}

@article{ClopperPearson1934,
  author  = {Clopper, C. J. and Pearson, E. S.},
  title   = {The Use of Confidence or Fiducial Limits Illustrated in the Case of the Binomial},
  journal = {Biometrika},
  year    = {1934},
  volume  = {26},
  number  = {4},
  pages   = {404--413},
  doi     = {10.1093/biomet/26.4.404}
}

@article{lacki2016,
  title={Type iii societies (apparently) do not exist},
  author={Lacki, Brian C},
  journal={arXiv preprint arXiv:1604.07844},
  year={2016}
}

@ARTICLE{Bradbury,
       author = {{Bradbury}, R.~J. and {Cirkovic}, M.~M. and {Dvorsky}, G.},
        title = "{Dysonian Approach to SETI: A Fruitful Middle Ground?}",
      journal = {Journal of the British Interplanetary Society},
         year = 2011,
        month = jan,
       volume = {64},
        pages = {156-165},
       adsurl = {https://ui.adsabs.harvard.edu/abs/2011JBIS...64..156B}
}

@article{Carrigan2009,
doi = {10.1088/0004-637X/698/2/2075},
url = {https://doi.org/10.1088/0004-637X/698/2/2075},
year = {2009},
month = {jun},
publisher = {The American Astronomical Society},
volume = {698},
number = {2},
pages = {2075},
author = {Carrigan, Richard A.},
title = {IRAS-BASED WHOLE-SKY UPPER LIMIT ON DYSON SPHERES},
journal = {The Astrophysical Journal}
}

@article{Prsa,
doi = {10.3847/0004-6256/152/2/41},
url = {https://doi.org/10.3847/0004-6256/152/2/41},
year = {2016},
month = {aug},
publisher = {The American Astronomical Society},
volume = {152},
number = {2},
pages = {41},
author = {Prša, Andrej and Harmanec, Petr and Torres, Guillermo and Mamajek, Eric and Asplund, Martin and Capitaine, Nicole and Christensen-Dalsgaard, Jørgen and Depagne, Éric and Haberreiter, Margit and Hekker, Saskia and Hilton, James and Kopp, Greg and Kostov, Veselin and Kurtz, Donald W. and Laskar, Jacques and Mason, Brian D. and Milone, Eugene F. and Montgomery, Michele and Richards, Mercedes and Schmutz, Werner and Schou, Jesper and Stewart, Susan G.},
title = {NOMINAL VALUES FOR SELECTED SOLAR AND PLANETARY QUANTITIES: IAU 2015 RESOLUTION B3*

†},
journal = {The Astronomical Journal}
}

@article{C,
title = {Galactic gradients, postbiological evolution and the apparent failure of SETI},
journal = {New Astronomy},
volume = {11},
number = {8},
pages = {628-639},
year = {2006},
issn = {1384-1076},
doi = {https://doi.org/10.1016/j.newast.2006.04.003},
url = {https://www.sciencedirect.com/science/article/pii/S1384107606000492},
author = {Milan M. Ćirković and Robert J. Bradbury}
}
\bibliographystyle{aasjournalv7}

\end{document}